\documentclass[12pt,a4paper,final]{iopart}

\usepackage{iopams}
\usepackage{cite}

\expandafter\let\csname equation*\endcsname\relax

\expandafter\let\csname endequation*\endcsname\relax

\usepackage{amsmath}
\usepackage{iopams}
\usepackage{graphicx}
\usepackage[breaklinks=true,colorlinks=true,linkcolor=blue,urlcolor=blue,citecolor=blue]{hyperref}

\usepackage{soul}

\begin{document}

\title[Binary lattice-gases of particles with soft exclusion]{Binary lattice-gases of particles with soft exclusion: Exact phase diagrams for tree-like lattices}	

\author{Dmytro Shapoval$^{1,2}$, Maxym Dudka$^{1,2,3}$,  Olivier B\'enichou$^4$ \& Gleb Oshanin$^4$}
\address{$^1$Institute for Condensed Matter Physics, National Academy of Sciences of Ukraine, 1 Svientsitskii Street, UA-79011 Lviv, Ukraine}
\address{$^2$ ${\mathbb L}^4$ Collaboration \& Doctoral College for the Statistical Physics of Complex Systems, Leipzig-Lorraine-Lviv-Coventry, Europe}
\address{$^3$Institute of Theoretical Physics, Faculty of Physics, University of Warsaw, Pasteura 5, 02-093 Warsaw, Poland}
\address{$^4$Sorbonne Universit\'e, CNRS, Laboratoire de Physique Th\'eorique de la Mati\`ere Condens\'ee (UMR CNRS 7600), 4 Place Jussieu, 75252 Paris Cedex 05, France}

\begin{abstract}
We study equilibrium
properties of binary lattice-gases comprising
$A$ and $B$ particles,
which undergo continuous exchanges with their respective reservoirs, maintained
at chemical potentials $\mu_A = \mu_B = \mu$. The particles
interact via on-site hard-core exclusion and also between the nearest-neighbours:
there are a soft \textit{repulsion} between $AB$ pairs and also 
interactions of arbitrary strength $J$, positive or negative,  for $AA$ and $BB$ pairs.
For tree-like  Bethe and Husimi lattices we determine the full phase diagram of such a ternary  mixture of particles and voids. 
We show that for $J$ being above a lattice-dependent threshold value,  
the critical behaviour is similar: the system undergoes a transition at $\mu = \mu_c$ from a phase with equal mean densities of species into a phase with a spontaneously broken symmetry, in which the mean densities are no longer equal.
Depending on the value of $J$, this transition can be either continuous or of the first order. 
For sufficiently large negative $J$, the behaviour on the two lattices becomes markedly different: 
on the Bethe lattice there exist  two separate phases with different kinds of structural order, which 
are  absent on the Husimi lattice, due to stronger frustration effects. 
\end{abstract}

\maketitle

%%%%%%%%%%%%%%%%%%%%%%%%%%%%%%%%%%%%%%%%%%%%%%%%%%%%%%%%%%%%%%%%
\section{Introduction}
%%%%%%%%%%%%%%%%%%%%%%%%%%%%%%%%%%%%%%%%%%%%%%%%%%%%%%%%%%%%%%%%
\setcounter{footnote}{0}

Lattice gases of particles with on-site and also
nearest-neighbour exclusion have been studied for a long time
as simple models exhibiting a transition from a disordered into an ordered phase.
Here, particles
adsorb onto the sites of a regular lattice  from a reservoir
subject to the constraint
that neither two particles can simultaneously occupy the same and the
neighbouring lattice sites, and may desorb spontaneously back to the reservoir.
A considerable knowledge of the thermodynamic properties of such systems has been gained through a series
of insightful analytical and numerical analyses.
Starting from the early works (see, e.g. Ref. \cite{domb} and references therein), different approaches have been proposed including, e.g.
  the Mayer cluster expansion \cite{cluster},
a variety of analyses based on the series expansions \cite{series1,series2,series3,series4,series5,series6},
the quasi-chemical and ring approximations
\cite{temperley},
the Bethe-lattice approximation \cite{bethe1,bethe2}  and so on,
 culminating at the exact solution obtained for the so-called "hard-hexagons" model
 \cite{baxter1,baxter2}.
More recently,  exact solutions have been found for  some random graphs and also for particles that differ in size \cite{bouttier}. An exact solution on the Bethe lattice has been derived for a model with  two kinds of particles -  smaller ones which occupy a single lattice site and larger ones which do not allow other particles to occupy its neighbouring sites \cite{tiago}.

In this paper we are concerned with a related class of equilibrium statistical mechanics models -- 
the so-called "reactive" lattice gases (RLG) -- in which chemical
 interactions between the neighbouring particles are interpreted as a
 nearest-neighbour exclusion constraint imposed on some kinds of particles. More specifically, there exists
a special type of chemical reactions that
take place whenever any two particles (similar or dissimilar, depending on the case at hand) encounter each other in a specified vicinity of a catalytic
 substrate. In such a situation, a reaction occurs either instantaneously or with some finite probability, and  both reactants disappear from the system  (see e.g. Refs. \cite{ziff,marro,evans,blumen,tox,argyrakis,cop} and references therein).
 In RLG models 
one considers 
a regular lattice of adsorbing sites, which is in contact with a reservoir of particles (or several reservoirs, in case when several types of particles are present) maintained at a constant chemical potential; the particles thus undergo continuous exchanges with their respective reservoirs - they adsorb onto vacant lattice sites and may spontaneously desorb from the lattice. It is supposed next
that either some of the lattice sites, or some of the bonds connecting neighbouring sites,
have a special catalytic property such that  
 the particles which enter into a "reaction"   
cannot appear simultaneously at the neighbouring lattice sites, whenever either of them or both are catalytic, or at the neighbouring sites connected by a catalytic bond. Otherwise, in absence of a catalyst, reactive particles coexist. This reactive constraint is then interpreted in such a way that, in equilibrium, configurations of particles  corresponding to a possible reaction event provide a zero contribution to the partition function, i.e. are forbidden. Note that for RLG with catalytic bonds a particle is linked by this bond to only one of its neighbours
and thus "interacts" only with it.  Conversely,
a particle residing on a catalytic site interacts with all of its nearest-neighbours,
which results in non-additive collective interactions. We also note that, despite some similarity, there is no one-by one correspondence between the RLG models and
the models of catalytic reactions introduced in Refs. \cite{ziff,marro,evans,blumen,tox,argyrakis,cop}. For the latter, the presence of an irreversible reaction drives the system out-of-equilibrium, while the RLG models are defined in equilibrium.

When  a lattice is completely covered by the catalytic bonds or sites,  a RLG becomes a gas of hard-core particles 
in which 
 the nearest-neighbour exclusion between the species which enter into a reaction is imposed on the entire lattice.  For a finite concentration of the catalyst, the situation is evidently more complicated and depends also on the way how the latter is distributed on the embedding lattice.
For both cases of catalytic bonds and catalytic sites present at arbitrary mean concentration,
with either \textit{annealed} or \textit{quenched} random spatial distributions of a catalyst,
exact solutions  have been obtained for one-dimensional lattices for
 single-species \cite{Oshanin2002,Oshanin2003,Oshanin2003a,Oshanin2003b} and for two-species RLG \cite{Shapoval2020}. In the former case, it was supposed that
 only one type of particles is present, while in the latter case it was assumed that there exist two dissimilar species, say $A$ and $B$, which have the same size (i.e. require a single vacant site for an adsorption) but differ in their chemical properties. Correspondingly,  in the two-species case a configuration in which  any $A$ and any $B$ appear at the neighbouring sites connected by either a catalytic bond, or when (at least) one of the occupied sites is in the catalytic state are forbidden, while similar species occurring at the neighbouring sites may coexist.
 For quenched disorder the models are solvable by combinatoric arguments or, alternatively, by expressing the adsorbate pressure as the Lyapunov index of an infinite product of random (uncorrelated for the model with catalytic bonds and sequentially correlated in case of catalytic sites)  
 matrices. In the \textit{annealed} disorder case such models reduce to those of lattice gases with a soft repulsion. One seeks then an appropriate recursion relation obeyed by the partition function and solves the latter by standard means.

For single-species RLG in higher dimensional systems the problem is clearly unsolvable in general and there
exists an exact solution only for a pseudo-lattice - the so-called Bethe lattice  with an \textit{annealed} disorder in placement of the catalytic bonds or sites \cite{Dudka2018}. In case of catalytic bonds,
the model is similar  to the one studied, e.g. in Ref. \cite{bethe2}, except for the fact that
here the repulsive interactions are not infinitely strong but are
"soft" and their amplitude depends on the mean concentration $p$ of the catalytic bonds; it diverges when $p \to 1$ only, but is finite for any $p < 1$.  It was shown that such a model exhibits a continuous transition from a disordered into an ordered phase at a certain value of the chemical potential; evidently,  for $p \to1$ this critical value converges to the one obtained in Ref. \cite{bethe2}. This transition is
followed  by a continuous re-entrant transition into a disordered phase (which is pushed to infinity when $p \to 1$, i.e. disappears in case of hard objects).
The case of catalytic sites is more complicated due to emerging multi-particle interactions and, in consequence, a critical behaviour is somewhat richer.
 While the direct transition into an ordered phase is always continuous, the re-entrant transition into the disordered phase may be either continuous or of the first order, depending on the value of the concentration $p_s$ of the catalytic sites.  The critical value of the chemical potential for the re-entrant transition is also pushed to infinity as $p_s \to 1$.

For RLGs with two kinds of species the only available exact solution has been obtained for  a
regular honeycomb lattice with a random \textit{annealed} distribution of the catalytic bonds in a rather special case:
the
chemical potentials of both species were taken equal to each other (such that both species are expected to be present, on average, at equal mean densities)  and with 
some  imposed restrictive condition that
the concentration $p$ of the catalytic bonds and the interaction
parameters are linked to each other through a certain relation.
It was shown \cite{Oshanin2004,Popesku2007}
that in this special case the model reduces  to an exactly solvable version of the Blume-Emery-Griffiths model, which maps onto the Ising model in a zero external field \cite{horiguchi}. The solution then predicts
 a non-trivial, fluctuation-induced continuous transition into a phase with a broken
 symmetry with respect to the mean densities of both kinds of particles. 
 It remains unclear, however, whether
 such a transition persists beyond this restrictive relation between the concentration and the interaction parameters,
 or it is a spurious phenomenon appearing  solely
 due to such a constraint.

 We revisit here the model considered in Refs. \cite{Oshanin2004,Popesku2007} from a broader perspective by
 relaxing the constraint imposed on $p$ and the interaction parameters.
 More specifically, we study here the thermodynamic properties of a two-species RLG with
  reactions between dissimilar species (i.e., with an imposed constraint that dissimilar species cannot appear simultaneously on neighbouring sites connected by a catalytic bond) in the completely symmetric case in which
    the chemical potentials of both species involved are equal to each other,
    as well as are the
 amplitudes $J$ of interactions between the neighbouring species of the same type. One expects, of course, that in this case
 both species in the binary RLG are present at equal, on average, densities. The mean concentration $p$, $0  \leq p \leq 1$, which defines the amplitude of the repulsive interactions between the dissimilar species, is an independent parameter,
 and the interaction amplitude $J$ between the neighbouring similar species
 is let to have an arbitrary magnitude and sign.
  We provide exact results for the thermodynamic properties of such binary RLGs
  on tree-like pseudo-lattices, as exemplified here by the Bethe lattice (see e.g. Ref. \cite{baxter2}) and the Husimi lattice (see e.g. Ref. \cite{harary}), with a random annealed distribution of the catalytic bonds.  We note that such an analysis corresponds to a certain mean-field-like approximation of the behaviour taking place on regular lattices; we remark, however,  that such an approach usually defines correctly the order of the phase transition, if any, and provides a rather accurate estimate of its location in the parameter space.
  We proceed to show that the symmetry
  breaking transition predicted in Refs. \cite{Oshanin2004,Popesku2007} is a valid feature of the model,
  but it is not the only
  phase transition taking place in such a system and moreover, this transition is not always continuous.
  We show that for an arbitrary $p$, and for $J$ exceeding a certain positive critical value $J_{tc}(p)$ (a tricritical point), the system enters
   into the phase with a broken symmetry  via the first order transition when the vapour pressure exceeds a certain critical value. For arbitrary $p$ and $J$ such that $J_{tc}(p) > J > J(p)$, where $J(p)$ is a  $p$-dependent threshold constant,
   the transition into the broken-symmetry phase is continuous. Such a behaviour is predicted for both tree-like lattices, and differs only in the precise location of the demarkation curve. Behaviour at negative values of $J$, when the similar species repel each other, is very different for the Bethe and the Husimi lattices. For the Bethe lattice we realise that
 in certain ranges of values of $J$ and $p$ there  exist two phases with a structural order: an alternating phase
 I in which the system spontaneously partitions into two sub-lattices - one being occupied by a mixture of both species present at equal mean densities, and the second one being  almost  devoid of particles, and  an alternating phase II in which each of the species resides predominantly on its own sub-lattice.
 The system enters into such phases and leaves one of them  
 via continuous phase transitions. On the contrary, such phases with structural order are absent on the Husimi lattice due to stronger frustration effects.

The paper is outlined as follows. In section \ref{model} we formulate our model, derive the corresponding effective
Hamiltonian and also describe the geometrical
features of the tree-like lattices under study.
The section \ref{bethe} is devoted to the analysis of the phase diagram of the RLG and the behaviour of the characteristic
 parameters for the Bethe lattice.
 In section \ref{husimi} our analysis is extended over the case of the Husimi lattice.
 Finally,  in section \ref{conc} we conclude with a brief
 summary of our results. Details of  intermediate derivations are relegated to Appendices.

\section{Model}
\label{model}

Consider an arbitrary lattice structure and suppose that some of the bonds connecting neighbouring sites $\langle ij\rangle$ possess a special "catalytic" property.   To specify the state of bonds, we associate with each of them a random variable
$\zeta_{\langle ij\rangle}$, such that  $\zeta_{\langle ij\rangle} = 1$ (with probability $p$)  if the bond
$\langle ij\rangle$ is catalytic and  $\zeta_{\langle ij\rangle} = 0$, otherwise, with probability $1-p$. Hence,  in the thermodynamic limit $p$ can be thought of as the mean concentration of the catalytic bonds.
Suppose next that the lattice is brought in contact
with two reservoirs of particles - $A$ and $B$ - which are identical in size but are distinguishable, e.g. by their colour. The reservoirs are kept at constant chemical potentials, in general, $\mu_A$ and $\mu_B$. The $A$ and $B$ particles undergo continuous exchanges with their respective reservoirs; they adsorb onto vacant lattice sites and may spontaneously   desorb. Particles of similar species occurring at neighbouring sites interact with each other and the strength of such interactions is denoted by $J_A$ or $J_B$, which can be negative or positive, i.e. we let the interactions be repulsive or attractive. Lastly, we stipulate that  the configurations in which an $A$ and a $B$ appear simultaneously at neighbouring  sites connected by a catalytic bond are forbidden.

Let $n_i$ and $m_i$ be two Boolean variables describing an instantaneous occupation of the site $i$. We use the convention that $n_i = 1$ ($m_i = 1$) if the site $i$ is occupied by an $A$ (a $B$) particle, and is zero, otherwise. Situations when both $m_i$ and $n_i$ are non-zero for the same site are not permitted because of the on-site exclusion.
In thermodynamic equilibrium and for a given realisation of random variables $\zeta_{\langle ij\rangle}$, the grand-canonical partition function of such a binary adsorbate, which counts the weights of different realisations of particles' placement on the lattice, writes
\begin{align}
\label{startZnCayley}
Z(\{\zeta_{\langle ij\rangle}\}) &= \sum_{\{n_{i},m_{i}\}} z_A^{\sum_{i}n_{i}} z_B^{\sum_{i}m_{i}} e^{\beta J_{A}\sum_{\langle ij\rangle}n_{i} n_{j} + \beta J_{B}\sum_{\langle ij\rangle}m_{i} m_{j}}   \nonumber \\ &\times  \prod_{\langle ij\rangle} (1-\zeta_{\langle ij\rangle} n_{i}m_{j})(1-\zeta_{\langle ij \rangle} m_{i}n_{j}),
\end{align}
where $\beta$  defines the reciprocal temperature measured in units of the Boltzmann constant,
while $z_A = \exp(\beta \mu_A)$ and $z_B = \exp(\beta \mu_B)$ are the activities, associated with the chemical potentials $\mu_A$ and $\mu_B$. Further on, in equation \eqref{startZnCayley},
the sum with the subscript $\{n_{i},m_{i}\}$
 signifies that the summation extends over all possible values of the occupation variables, while the sums (as well as the product) with the subscript $\langle ij \rangle$ denote the summation (the product) extending
 over all bonds connecting nearest-neighbouring lattice sites.
  The factor in the second line in equation \eqref{startZnCayley} forbids
 the configurations in which an $A$ and a $B$ appear simultaneously at the neighbouring sites connected by a catalytic bond such that their contribution to the partition function equals zero.

Suppose next that the disorder in the placement of the catalytic bonds is annealed and hence, the grand-canonical partition function in equation \eqref{startZnCayley} can be directly averaged over random variables $\zeta_{\langle ij \rangle}$. Because the variable
$\zeta_{\langle ij\rangle}$ assigned to a given bond is statistically independent of the state of other bonds, such an averaging is straightforward and
yields the following form of the grand-canonical partition function
\begin{align}
\label{averagedZnCayley}
Z(p) &= \sum_{\{n_{i},m_{i}\}} z_A^{\sum_{i}n_{i}} z_B^{\sum_{i}m_{i}} e^{-\beta {\cal H}} ,
\end{align}
where ${\cal H}$ is the effective Hamiltonian of the model under study :
\begin{align}
{\cal H} = - J_{A}\sum_{\langle ij\rangle}n_{i} n_{j} - J_{B}\sum_{\langle ij\rangle}m_{i} m_{j} - \beta^{-1} \ln(1-p) \sum_{\langle ij\rangle}(m_{i} n_{j} + n_{i} m_{j}) \,.
\end{align}
Note that upon  averaging over the catalytic properties of the bonds, a strict exclusion constraint gets replaced by a  milder condition: in case of an annealed disorder in placement of the catalytic bonds the dissimilar species
exhibit a short-range mutual repulsion with a finite amplitude $-\ln(1-p)$, which becomes infinitely strong for $p=1$ only.
Hence, for $p < 1$, 
an $A$ and a $B$
may, in principle,
reside simultaneously on the neighbouring sites but
 there is a penalty to pay.
In what follows we will concentrate exclusively on the symmetric case with $\mu_A = \mu_B = \mu$ and $J_A = J_B = J$, such that one may expect that $A$s and $B$s are present in the system at equal mean densities. With such parameters our system maps to some kind of Blume-Emery-Griffits \cite{BEG} model, known,  as well as its generalizations, to describe systems with tricritical behaviour (see e.g. Refs~\cite{Mukamel74,Prasad2019} and reference therein)

\begin{figure}%[htbp]
	\begin{center}
		\includegraphics[width=0.38\hsize]{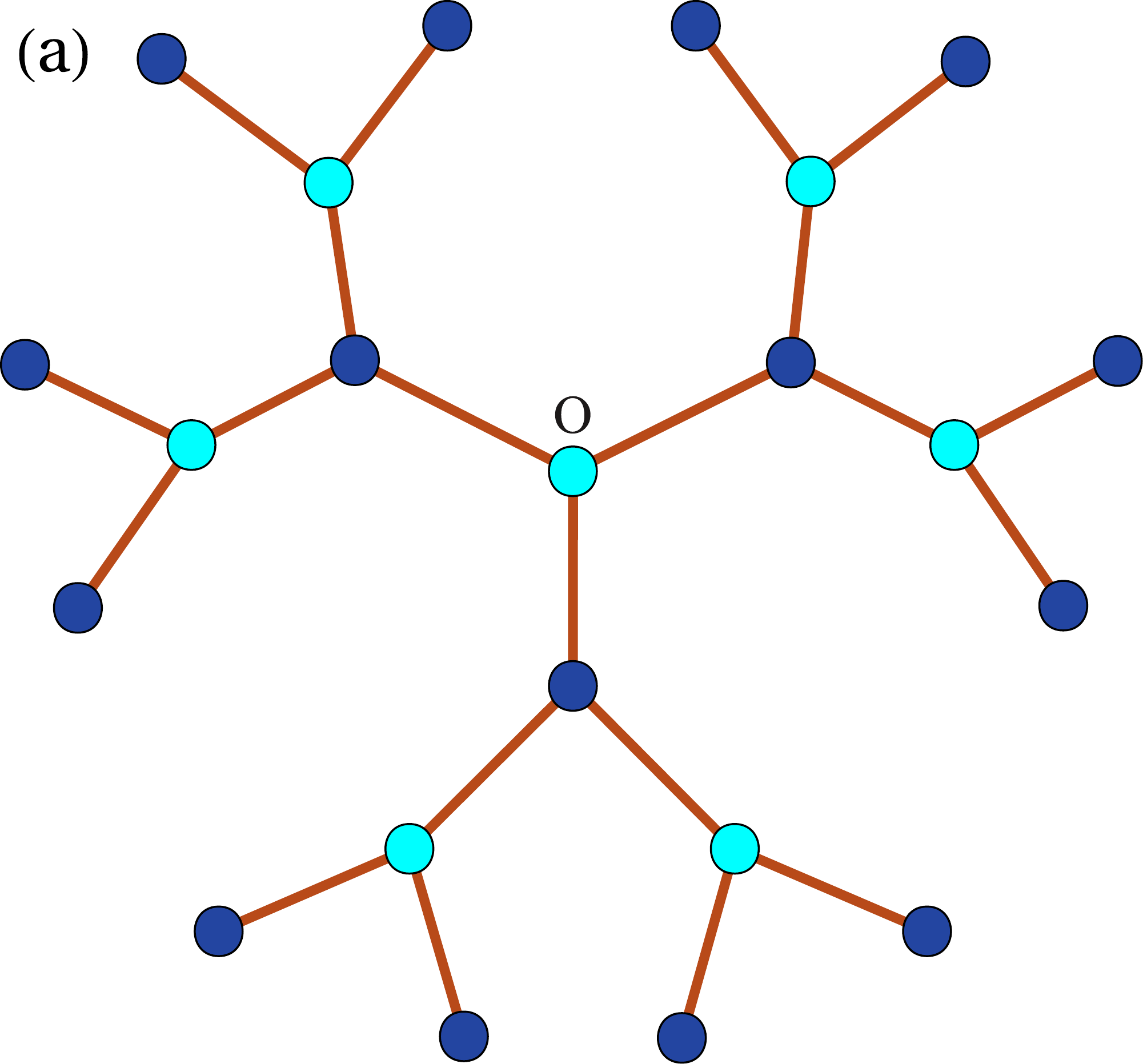} \hfill
		\includegraphics[width=.42\hsize]{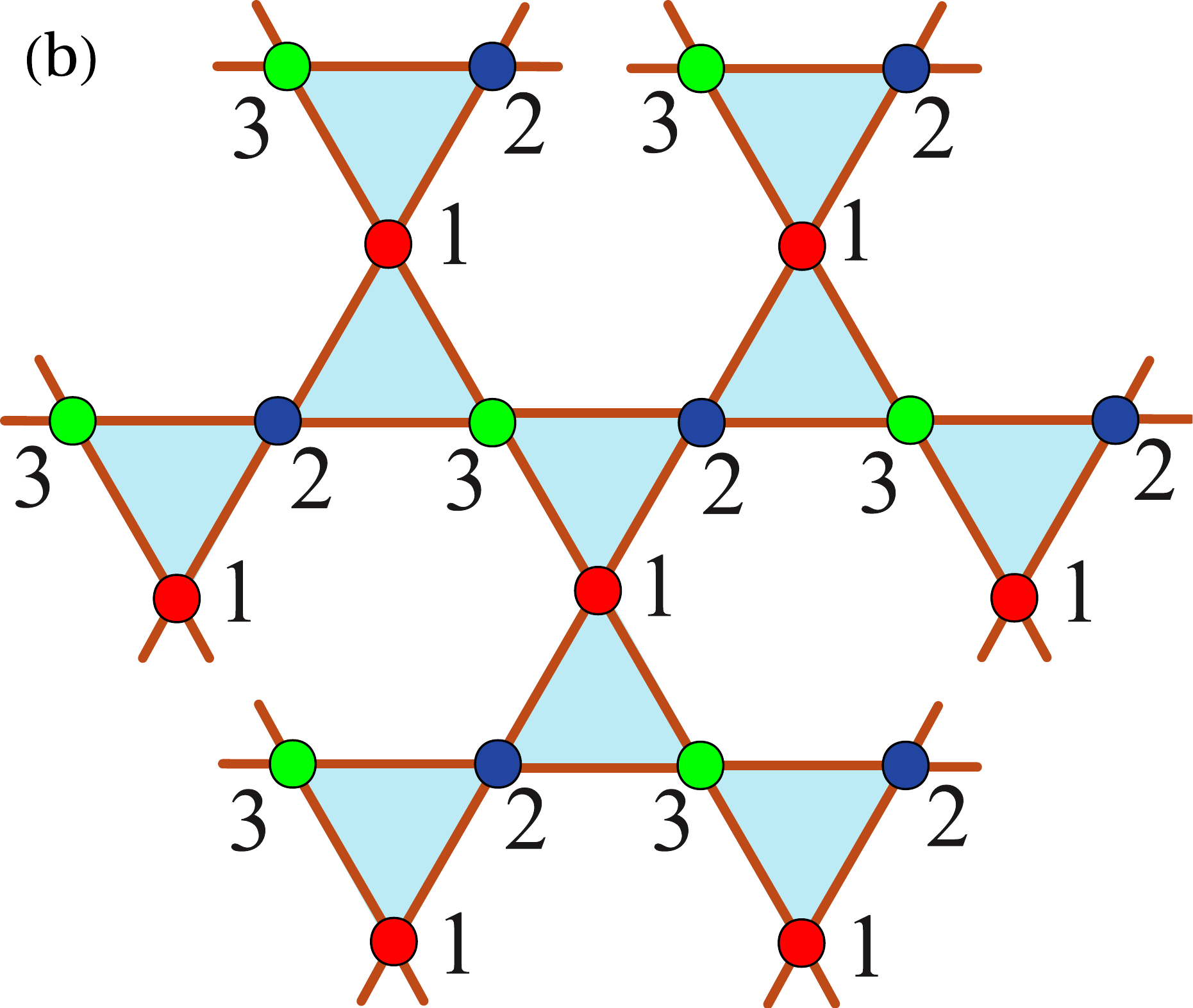}
	\end{center}
	\caption{Panel (a). The Cayley tree with the coordination number $q=3$ containing $N = 3$ generations, emanating from the root site $O$. Such a Cayley tree is bi-partite and two sub-lattices 	are marked by different colours. Panel (b). A fragment of the Husimi lattice  with the elementary units being the triangles and the coordination number of triangles $t = 2$.  Such a lattice is three-partite; three different sub-lattices are denoted by $1$, $2$, and $3$, respectively.}
	\label{fig1}
\end{figure}

Our analysis of the partition function in equation \eqref{averagedZnCayley} is performed for two particular cases of pseudo-lattices -
 the Bethe lattice and the Husimi lattice (see Fig. \ref{fig1}) \cite{harary}.
 
The Bethe lattice represents
a deep interior (far away of the boundary sites) of the so-called
Cayley tree with an infinite number of generations (see  Fig. \ref{fig1}, panel (a), for an example of the Cayley tree with three generations). In turn, the Husimi lattice (see  Fig. \ref{fig1}, panel (b)) in the general case represents a deep interior part  with all equivalent vertices of
 a connected graph whose building blocks
  are  $Q$-polygons ($Q \ge 2$).  In Fig. \ref{fig1}, panel (b), we depict the Husimi lattice used in our analysis for which
  the
   polygons are triangles, i.e. $Q = 3$, and the coordination number $t = 2$ is the number of triangles which meet at each site. Hence, the Husimi lattice under study is an infinitely nested set of corner-sharing triangles such that
   the local geometry is identical to that of the kagome lattice. It has, however,
   a weaker geometrical frustration than the original  kagome lattice, because the triangles never reconnect resulting in a tree-like structure.
 More details about such tree-like lattices can be found in Refs. \cite{baxter2,harary,pre}.

We close this section by mentioning that such tree-like lattices have been often used for the analysis
of various models of statistical mechanics (and beyond). It was also understood in the past  (see, e.g. Ref. \cite{bethe2})
that
different approximate approaches, such as, e.g. the Bethe-Peierls approach, or some other approximations mentioned
in the Introduction, correspond in fact to a replacement of the actual regular lattice by some pseudo-lattice.  At the same time,
the analysis performed on a pseudo-lattice provides
a reliable approximation, and often a substantial improvement as compared to the mean-field calculations.
The Bethe lattice approach has been employed, among many other examples, for the analysis of
the  phase diagram of the Blume-Emery-Griffiths model \cite{Ananikian1991},
of modulated phases emerging in the Ising model with competing interactions~\cite{Vannimenus1981},
the Potts model~\cite{Ananikian2013},
lattice models of glassy systems~\cite{Rivoire2004} and localisation transitions~\cite{DeLuca2014}, as well as a phase behaviour of confined ionic liquids
\cite{Dudka2016,Dudka2019}.  Bethe lattice approach was also used in the  analytical studies of
different diffusion-reaction processes
\cite{Majumdar93,Abad04,Shapoval2018},
processes of random and cooperative sequential adsorption \cite{Chatelain12} or of the structural properties of branched polymers \cite{Henkel96,Rios01}.
The accuracy of such an approximation has been recently assessed in the numerical analysis of the Blume-Capel model and it was shown that, somewhat surprisingly,  it provides a very accurate estimate of the location of the demarkation curve between the ordered and disordered phases  \cite{groda}. On the other hand,
the Bethe-lattice approach cannot fully
reproduce the geometrical frustration in case when some competitive interactions are present, (e.g. when the coupling parameter $J$ in our model is negative and sufficiently large by its absolute value, while $z$ is sufficiently large prompting the system to accommodate more particles than the interactions permit).
To incorporate the frustration effects and thus to describe the system more adequately, one devises more elaborate
 cluster versions in which each vertex is replaced by a frustrated geometrical unit. When the latter is taken to be
  a triangle, the resulting structure is precisely the Husimi tree which is considered here (see Fig.~\ref{fig1}). 
  Analyses of different physical models on such pseudo-lattices are quite ubiquitous and appear in various physical contexts. We just mention recent studies of 
 antiferromagnetic classical   \cite{Jurcisinova2016,Jabar2018} and quantum spin models  \cite{Liao2016}, and  also several multi-site interaction models \cite{Jurcisinova2014c}. We finally remark that although such approaches usually
 quite accurately predict the location of the demarkation curves between ordered and disordered phase, as well as the order
 of the phase transition, they naturally fail to provide correct values of the critical exponents. As a matter of fact, both tree-like structures  can be considered as "infinitely-dimensional" systems\footnote{ Effective dimension of a system can be defined as the ratio of logarithms of the volume and of the linear extent in the limit when the latter tends to infinity. In doing so, e.g., for regular two-dimensional and three-dimensional lattices one finds the values $2$ and $3$, respectively. Following Ref. \cite{baxter2},  for the Bethe or the Husimi lattices the effective dimension 
is defined as the ratio of a logarithm of the number of sites in a tree with $N$ generations and $\ln N$. Evidently, 
this ratio tends to infinity when $N \to \infty$.}, and as a consequence, they  are characterised by the so-called "mean-field" values of the critical exponents.

%%%%%%%%%%%%%%%%%%%%%%%%%%%%%%%%%%%%%%%%%%%%%%%%%%%%%%%%%%%%%%%%
\section{The Bethe lattice}
\label{bethe}
%%%%%%%%%%%%%%%%%%%%%%%%%%%%%%%%%%%%%%%%%%%%%%%%%%%%%%%%%%%%%%%%

As we have remarked above, virtually all known classical
models of
statistical mechanics have been studied
on the Bethe lattice such that a general procedure of the derivations
of the partition function is well-documented (see, e.g. Refs. \cite{bethe2,baxter2,Dudka2018}). 
For the case at hand, which has some interesting peculiar features,  
we merely present below a brief summary of the steps involved and relegate the details to the \ref{A}.

Consider a Cayley tree with an
arbitrary coordination number $q$ and an arbitrary number $N$ of generations (see  Fig.~\ref{fig1} for an example with $q =3$ and $N=3$).  Upon specifying the occupation of the root site $O$, the Cayley tree naturally decomposes into $q$ rooted trees, such that $Z(p)$ in equation \eqref{averagedZnCayley} can be formally written down as
\begin{eqnarray}
\label{ZCayley}
Z (p) = C_{N}^{q}(0,p) +  z^{1-q} C_{N}^{q}(A,p) + z^{1-q} C_{N}^{q}(B,p),
\end{eqnarray}
where
$ C_{N}(0, p)$, $C_{N}(A, p)$ and $C_{N}(B, p)$ are the grand canonical partition functions of each rooted tree  conditioned on the occupation of
 the root site $O$: here, the  arguments $0$, $A$ and $B$ indicate that the root site
is vacant, occupied by an $A$, or by a $B$ particle, respectively.
 In turn, each rooted tree with $N$ generations consists of $q - 1$ identical subtrees with $N-1$ generations, and so on,  
such that a general strategy is to express the functions $C_N$ corresponding 
to a tree with $N$ generations,  via the functions $C_{N-1}$ corresponding to a tree with $N-1$ generations. 
A derivation 
of the recursion relations obeyed by functions $C_N$ in the general case is presented in \ref{A}. In the symmetric case, the latter can be conveniently  
written 
in terms of two auxiliary variables
\begin{eqnarray}
\label{newvar2}
x_{N} = \frac{C_{N}(A,p)}{z C_{N}(0,p)}, \qquad {\text{and}} \qquad y_{N} = \frac{C_{N}(B,p)}{z C_{N}(0,p)},
\end{eqnarray}
that obey the coupled non-linear recursion scheme of the form
\begin{eqnarray}
\label{FinalCayleyRec2}
x_{N} = \frac{1 + z e^{\beta J} x_{N-1}^{q-1} + (1-p) z y_{N-1}^{q-1}}{1 + z x_{N-1}^{q-1} + z y_{N-1}^{q-1}}, \nonumber \\
y_{N} = \frac{1 + (1-p) z x_{N-1}^{q-1} + z e^{\beta J} y_{N-1}^{q-1}}{1 + z x_{N-1}^{q-1} + z y_{N-1}^{q-1}}.
\end{eqnarray}
 We turn next to the limit $N \to\infty$ and consider a fragment of the Cayley tree which is in a deep interior part of a system well away from the boundary, i.e., which forms the so-called Bethe lattice \cite{baxter2}. The sites  of the latter
are all equivalent and hence, the thermodynamical phases are described by fixed point (or cycle solutions) $\{x, y\}$ of equation~(\ref{FinalCayleyRec2}), i.e., all $\{x_N, y_N\}$ should
converge to $\{x, y\}$ as $N \to \infty$. Given $\{x, y\}$, one determines the ensuing thermodynamic properties of our model. In particular, the mean densities of the $A$ and $B$ species on any site of the Bethe lattice are the same, 
and can be expressed via the
  grand partition functions $ C_{N}(0, p)$, $C_{N}(A, p)$ and $C_{N}(B, p)$  (see \ref{DER}) to give
  \begin{eqnarray}
\label{meandens4}
\rho^{(A)} = \frac{z x^{q}}{1 + z  x^{q} + z y^{q}}, \qquad
\rho^{(B)} = \frac{z y^{q}}{1 + z x^{q} + z y^{q}}.
\end{eqnarray}
Note finally that the derivation 
of the correct expression for the
free energy requires some additional and rather subtle arguments, because the contributions due to the boundary sites have to be properly excluded. This procedure is described in~\ref{Bfree}.

The subsequent analysis focuses on stability of the attractors of coupled recursions \eqref{FinalCayleyRec2}
for different values of the parameters $p$, $J$ and $z$. Here, four different situations are encountered:\\
-- In some region in the parameter space
two sequences $x_N$ and $y_N$  converge to a unique value $x = y$ as $N \to \infty$, which corresponds to a disordered phase with equal mean densities of $A$ and $B$ particles. In what follows, we coin such a phase as disordered symmetric phase.\\
--
 There is a domain in the parameter space in which such a convergence does not take place and instead, $x_N$ and $y_N$ converge to different limiting values $x$ and $y$ in the limit $N \to \infty$. This  means, in virtue of equations \eqref{meandens4},  that here
the symmetry between $A$s and $B$s is spontaneously broken and the mean densities of the components
become different; one of them ($A$ or $B$ with equal probability) is present in majority with a higher mean density  $\rho^{(+)}$, while the other one appears in minority with the mean density $\rho^{(-)}$, $\rho^{(+)} > \rho^{(-)}$. We call such a phase  -- the phase with a broken symmetry (PBS).\\
%To describe both latter situations, one first supposes that the tree is infinitely large, $N \to \infty$, and restricts the analysis to the sites which appear in the deep interior of such a tree, well away from its boundary. Here, all sites (including the root site $O$)  are equivalent and hence,  the values of the mean densities are independent of $N$. \\
-- There is a third situation in which
$x_N$ ($y_N$) with odd $N$ converges to one value $x_{odd}$ ($y_{odd} = x_{odd}$), while for even $N$ it ultimately tends to a different value $x_{even}$ ($y_{even} = x_{even}$).  Such a kind of the convergence, (the so-called subsequence convergence), is known to emerge in diverse models of statistical mechanics defined on recursive lattices. It is well-understood that, in fact, it is a manifestation of a spontaneous ordering phenomenon, i.e.,
an alternating partition of the particle phase, such that $A$s and $B$s occupy predominantly  just one of two sub-lattices,  while the second sub-lattice is almost empty. 
In this phase the mean densities of  both species are equal to each other but there is a structural order. In what follows, we refer to such a phase as a symmetric phase with an alternating order I (PAO I). \\
-- For some values of the parameters  a fourth situation is realised in which
$x_N$  with odd $N$   and $y_N$ with even $N$ converge to one value $x_{odd}=y_{even} $, while $x_N$  for even $N$ and $y_N$  for odd $N$ tend to a different value $x_{even}=y_{odd}$. Physically, it corresponds to a situation, in which $A$s  occupy predominantly one of the sub-lattices, while $B$s reside on the second one. Here, the mean densities of both species are also equal to each other. We refer to such a phase as a symmetric phase with an alternating order II (PAO II).\\
A salient feature of two last situations is that there are corresponding phase transitions: at entering the PAO I from a symmetric disordered phase, and
while leaving it to (or re-entering) a symmetric disordered phase. Similarly, there is a continuous transition upon entering the PAO II.
Here,  to tackle analytically  an emerging subsequence convergence, one reiterates the recursions \eqref{FinalCayleyRec2} expressing $x_N$ and $y_N$ through $x_{N-2}$ and $y_{N-2}$, (instead of $x_{N-1}$ and $y_{N-1}$), such that $N$ and $N-2$ appear to have the same parity. Then, the analysis proceeds in the same way as for the symmetric disordered phase or the phase with a broken symmetry, i.e. one goes to the limit $N \to \infty$ and concentrates on the sites which belong to the Bethe lattice. The order of the transitions which take place while crossing the demarkation line between different phases is obtained in a standard way from the analysis of the behaviour of the free energy (see equation \eqref{FreeEnergy}) and of the thermodynamic properties at the transition points.

\begin{figure}%[htbp]
	\begin{center}
		\includegraphics[width=0.47\hsize]{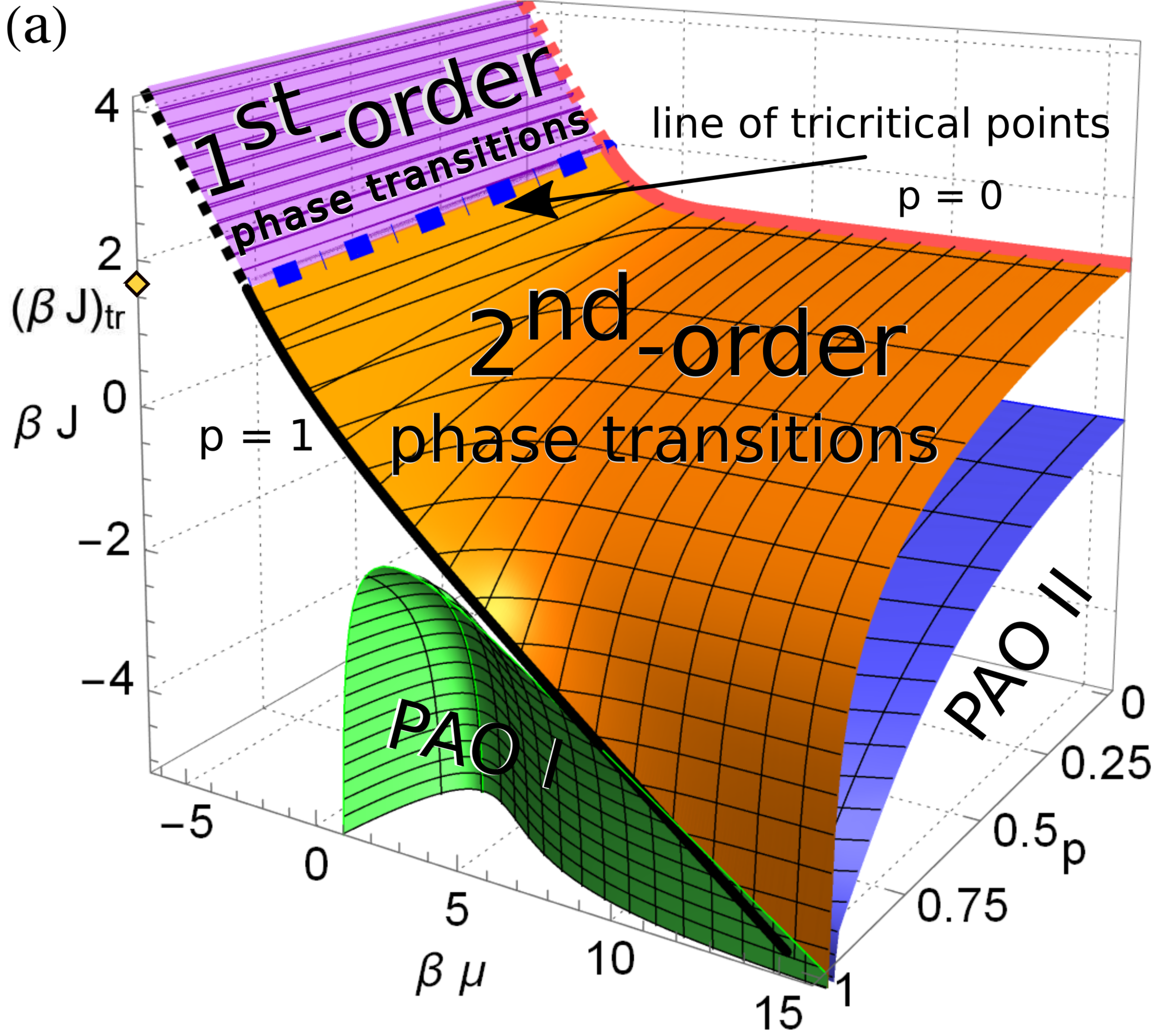} \hfill
		\includegraphics[width=0.45\hsize]{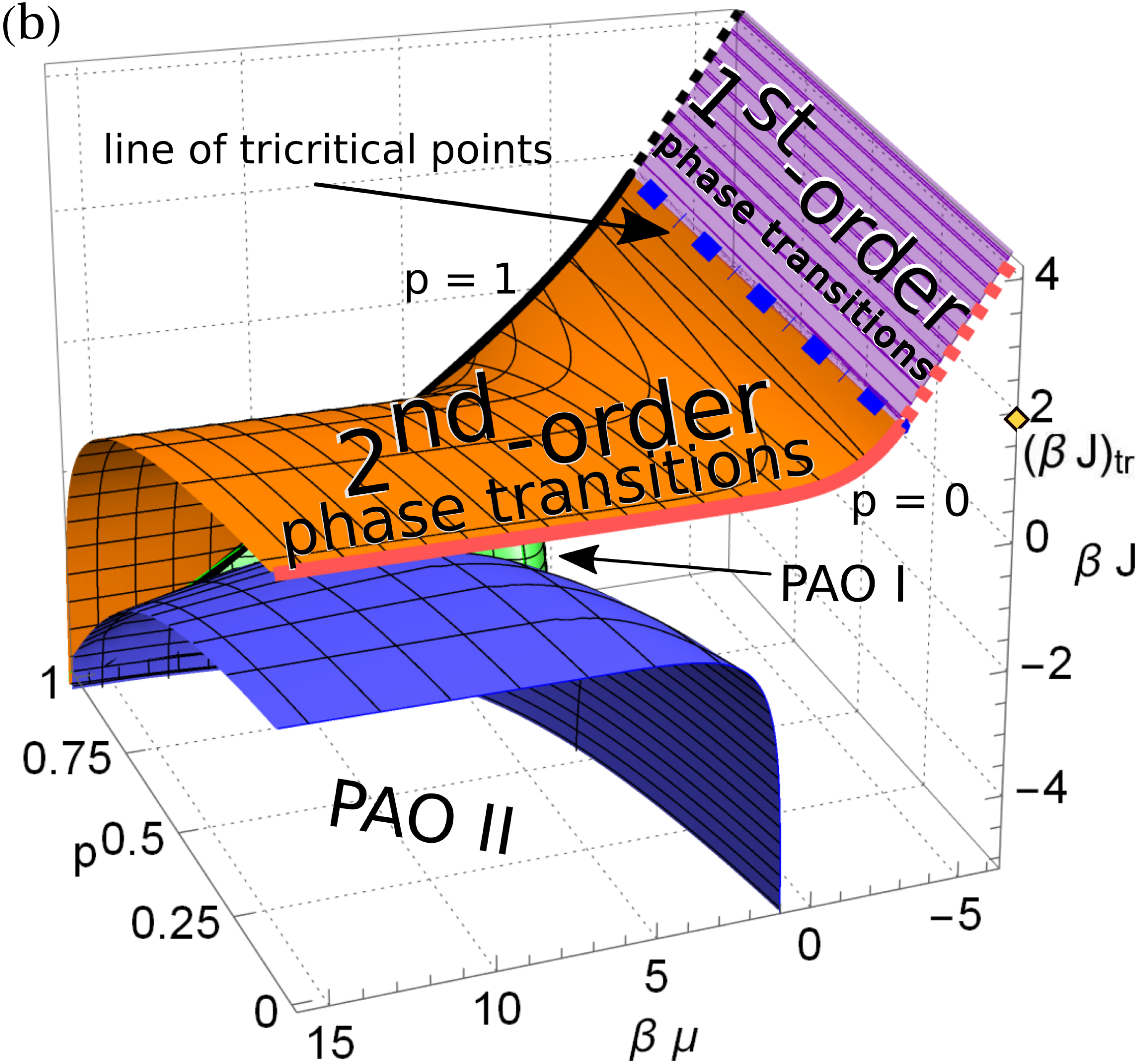} 
		\includegraphics[width=0.47\hsize]{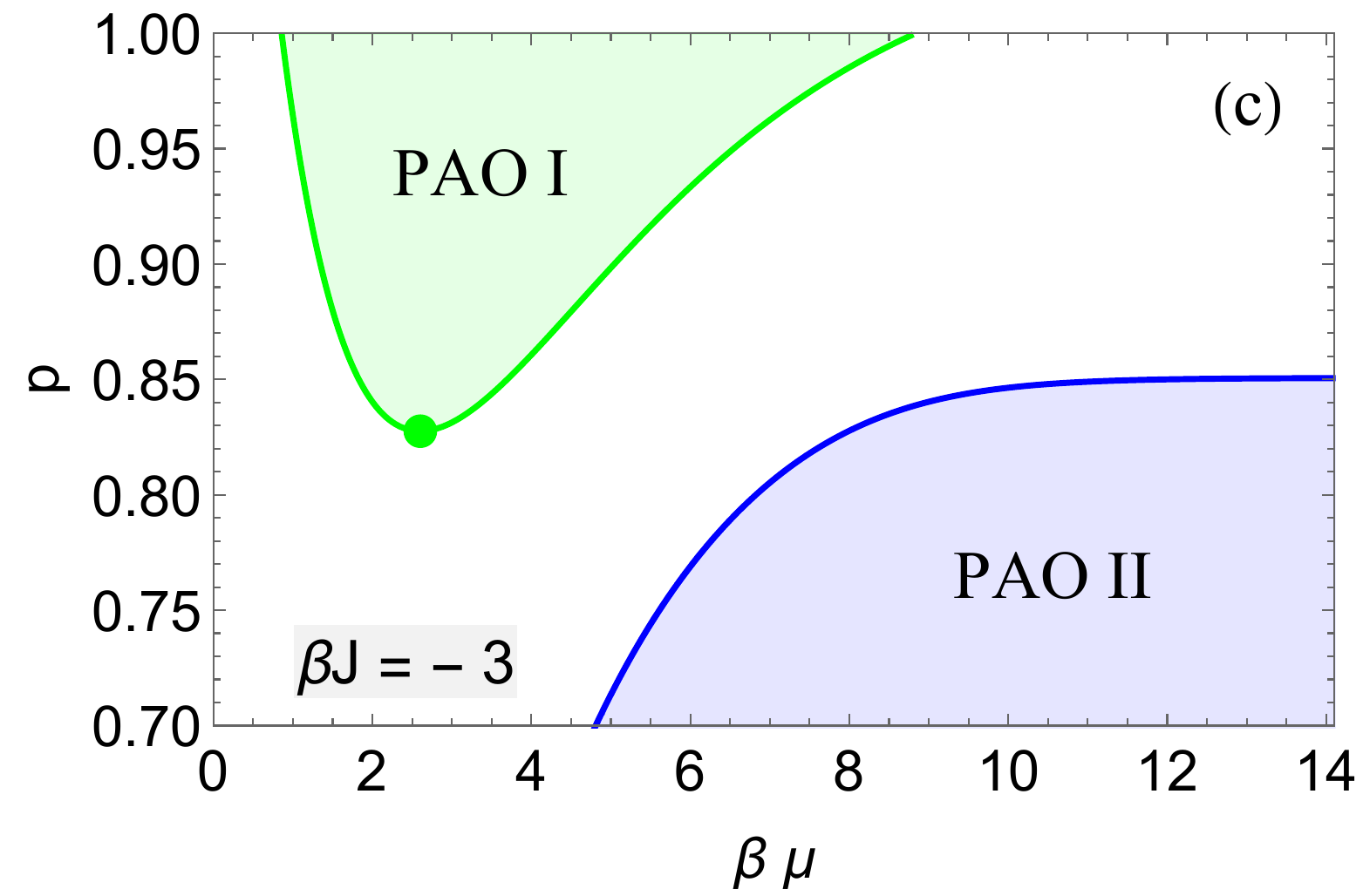}  \hfill
		\includegraphics[width=0.48\hsize]{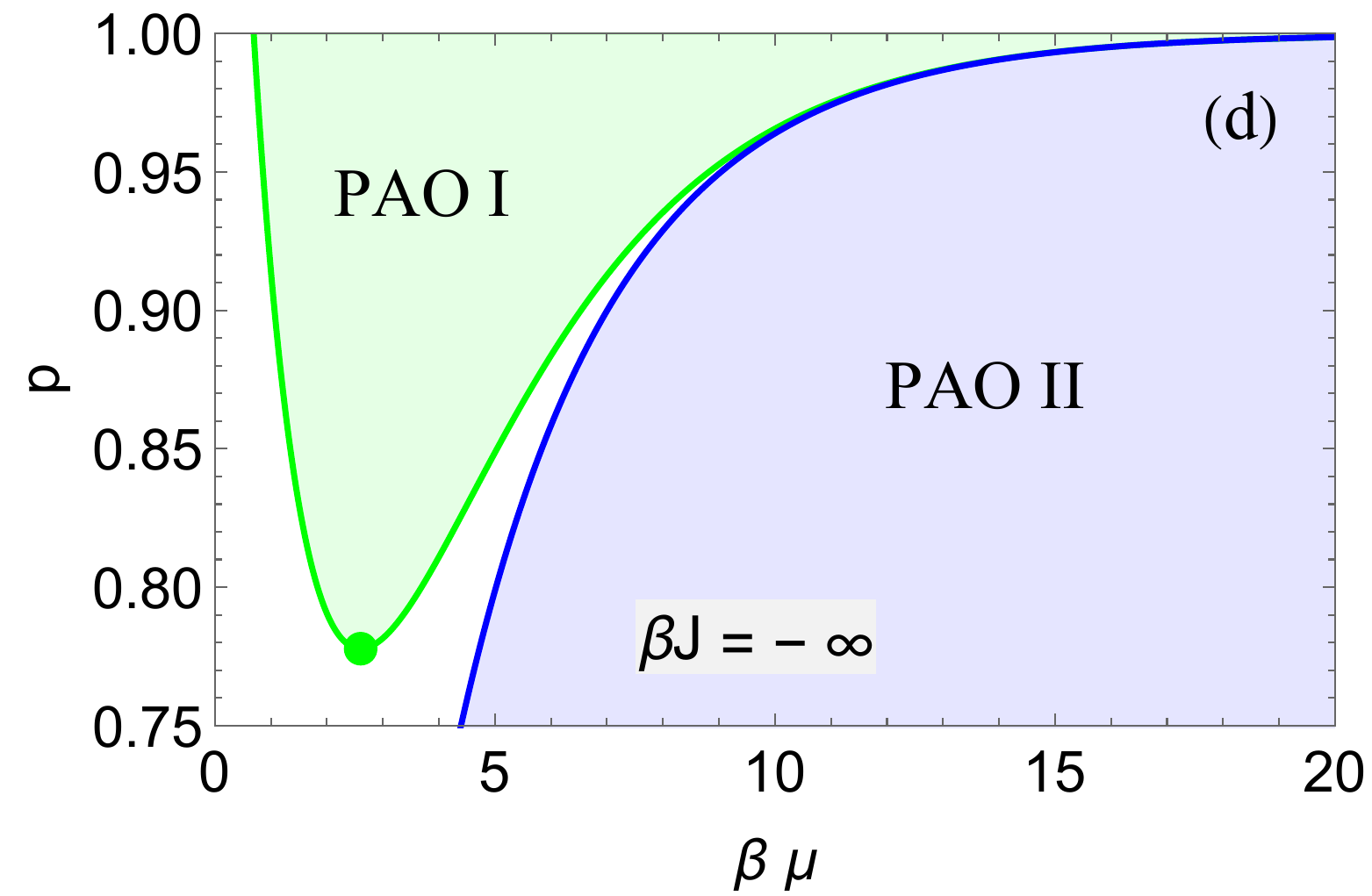}
	\end{center}
	\caption{(Colour online) Panel (a) : Phase diagram in the parameter space ($J,\mu,p$) for the Bethe lattice with the coordination number $q=3$.  The shaded surface divides the parameter space into two sub-spaces: the one above the surface corresponds to a phase with a broken symmetry (PBS) (see also panels (b) in Figs. \ref{rec-rel1} and \ref{fig3}), 
while
	below one has a symmetric phase with equal $A$ and $B$ particles densities. 
	The thick red and thick black curves indicate  the crossings of the demarkation surface with the $p=0$ and $p=1$- planes.
	A part  of the shaded surface (purple) above the thick dash-dotted curve (blue) -- the line of tricritical points -- corresponds to the first order transition into the PBS, while below this curve (a part of the surface painted by orange colour) the transition is continuous ($2^{\rm nd}$ order). 
	  Shaded surfaces within the symmetric disordered phase  
	  bound phases with an alternating order (PAO):  the PAO I, in which  $A$s and $B$s both occupy predominantly either of the sub-lattices, the second one being almost empty, is inside the green one,
	  while the blue one envelopes  the PAO II, in which the 
	  two species separate residing predominantly on two different sub-lattices. The panel (b) depicts the phase diagram on the panel (a) rotated on $90^{\circ}$, to make apparent the PAO II.
Panels (c) and (d) present the phase diagrams on the $(p,\mu)$-plane for two fixed values of $J$: $\beta J = - 3$ (c)
and  $\beta J = - \infty$ (d), i.e., for an infinitely strong repulsion between similar species.
Filled (green) circles denote the critical value $p_{c,1}$ of the parameter $p$, below which the PAO I does not exist; 
 $p_{c,1} \approx 0.8276$ (c) and $p_{c,1} = 7/9$ (d). For $p \geq p_{c,1} $, the PAO I exists for $\beta \mu$ within the interval
 $(\beta \mu)_{c, 1} \leq \beta \mu \leq (\beta \mu)_{c, 2}$,  equations \eqref{zc1}) and  \eqref{zc2}.
 The PAO II 
 occurs for $\beta \mu > (\beta \mu)_{c, 3}$, equation \eqref{zc3}, and $p \leq p_{c,2} $. In particular, 
  $p_{c,2} \approx 0.8506$ for $\beta J = - 3$ (c) and  $p_{c,2} = 1$ 
 for $\beta J = - \infty$ (d).}
	\label{ph-dim}
\end{figure}

The analysis of solutions of coupled recursion schemes \eqref{FinalCayleyRec2} permits us to
 construct the full phase diagram of our model on the Bethe lattice.
We depict it in Fig. \ref{ph-dim} for the particular case $q=3$. More general results for arbitrary $q \geq 3$ and details of calculations are presented in  \ref{A}. Note that the results for $q = 3$ and $q > 3$ are qualitatively similar and differ only in the precise values of the parameters at which a critical behaviour takes place.
The phase diagram for $q = 3$ is presented in the space of three parameters:  interaction strength $J$, chemical potential $\mu$ and the mean density $p$ of the catalytic bonds, which controls the amplitude of repulsive interactions between neighbouring dissimilar species. The phase diagram shows that a binary lattice-gas of particles with such interactions can be either in a symmetric phase, in which the mean densities of $A$ and $B$ particles are equal to each other, or in a phase in which such a symmetry is broken - the PBS, in which the mean densities of both species are no longer equal. Moreover, our analysis reveals that the symmetric phase itself divides into three sub-phases: a disordered symmetric phase and symmetric phases with two types of structural alternating order (PAO I and PAO II), the properties of which will be discussed at the end of this section.

\subsection{Symmetric phase versus the phase with a broken symmetry (PBS).}
\label{qq}

The symmetric phase and the PBS are separated by a demarkation surface defined as an implicit solution
of equations \eqref{dis1a} and \eqref{ord1.1a}, which are presented in \ref{A}. The PBS is situated above the demarkation surface, while the symmetric phase in which the mean densities of the $A$ and $B $ particles are equal to each other is below this surface.

On this surface,  there exists a line of tricritical points defined by
\begin{eqnarray}
\label{jtr}
(\beta J)_{tc}=\ln \left(4 - p + {\sqrt{9 -4 p (1 - p)}} \right) \,.
\end{eqnarray}
The value of $\beta J$ at the tricritical point is thus a slowly (logarithmically) varying function of the mean density of the catalytic bonds; in particular, for $p =0$, (such that there is no repulsion between neighbouring $A$s and $B$s),   we have $(\beta J)_{tc} = \ln (7)\simeq 1.946 $. For  $p =1$, (such that $A$s and $B$s are not allowed to occupy the neighbouring sites), $(\beta J)_{tc} = \ln (6) \simeq 1.792 $. For intermediate values of $p$, the value of the interaction strength  $(\beta J)_{tc}$ at the tricrital point smoothly interpolates between these two (not very different) numbers. Note that $(\beta J)_{tc}$ is always positive, such that the tricritical points exist only in case of
attractive interactions between similar species.

Suppose next that we fix $\beta \mu$ (or the activity $z$) and $p$, and vary $\beta J$ from some large negative value to a  positive one, such that for a certain value of $\beta J$ we cross the demarkation surface. This critical value, i.e. $(\beta J)_c$ corresponds to a transition point from a phase with equal densities of the $A$ and $B$ particles to the PBS. If such a crossing occurs above the line of tricritical points, the transition is of the first order and manifests itself via a discontinuity in the values of densities, while a crossing below this line corresponds to a continuous ($2^{\rm nd}$ order) transition. Similarly, if we fix $p$ and $\beta J > J(p)$, where $J(p)$ is some $p$-dependent  threshold value of the strength of interactions between similar species:
\begin{equation}
\label{Jp}
\beta J(p) = \ln\left(3 (1-p)\right) \,,
\end{equation}
and increase $\beta \mu$ from some large negative value to a sufficiently large positive value, we make the system to undergo a phase transition into the PBS. The order of a transition depends on the values of $p$ and $\beta J$ :
 if $\beta J > (\beta J)_{tc}$, equation \eqref{jtr}, the transition is of the first order, while for $\beta J(p) < \beta J <  (\beta J)_{tc}$ the transition is continuous and happens at the value of the activity which is equal to
 \begin{eqnarray}
\label{criticalz}
z_c = \exp\left((\beta \mu)_c\right) =  \frac{(e^{\beta J} + 3p -1)^2}{4 (e^{\beta J} - 3 (1-p))(e^{\beta J} - 1+p)^{2}}.
\end{eqnarray}
For $\beta J < \beta J(p)$, equation \eqref{Jp}, (which inequality can also be re-interpreted as some restriction imposed on the value of $p$), no such transition takes place and the system remains in the symmetric phase for any value of $\beta \mu$. It follows from equation \eqref{criticalz} that for $p = 0$ the critical value of the activity is simply $z_c = 1/(4 (\exp(\beta J)  - 3))$, (see the thick red curve in Fig. \ref{ph-dim}), and hence, no transition takes place for such a value of $p$ for $\beta J < \ln(3)$. In the opposite limit, i.e., for $p \to  1$, (see the thick black curve  in Fig. \ref{ph-dim}), $J(p) \to - \infty$ and hence, a transition exists for any sign and value of $J$.
	\begin{figure}%[htbp]
		\begin{center}
			\includegraphics[width=0.47\hsize]{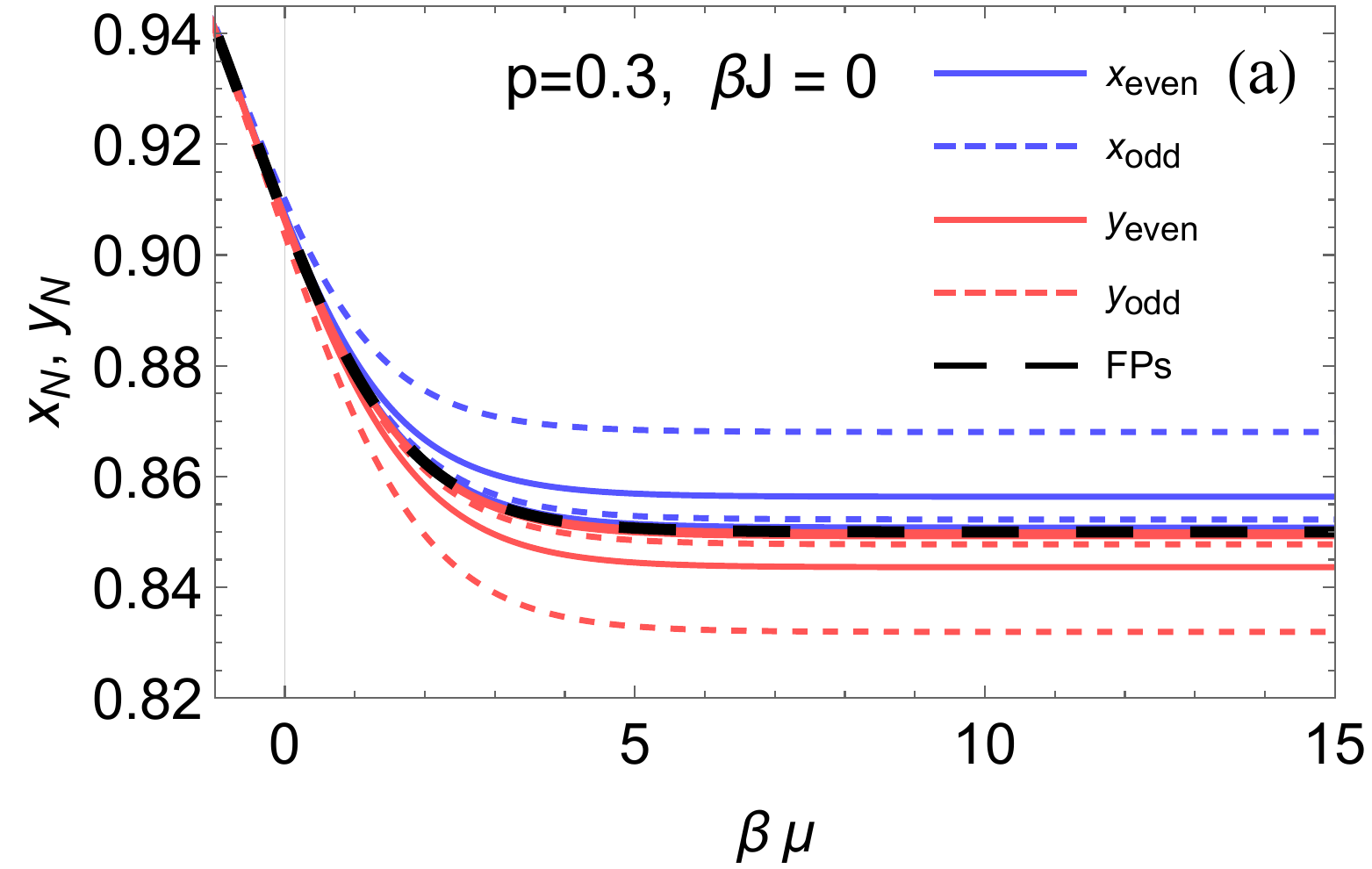} \hfill
			\includegraphics[width=0.47\hsize]{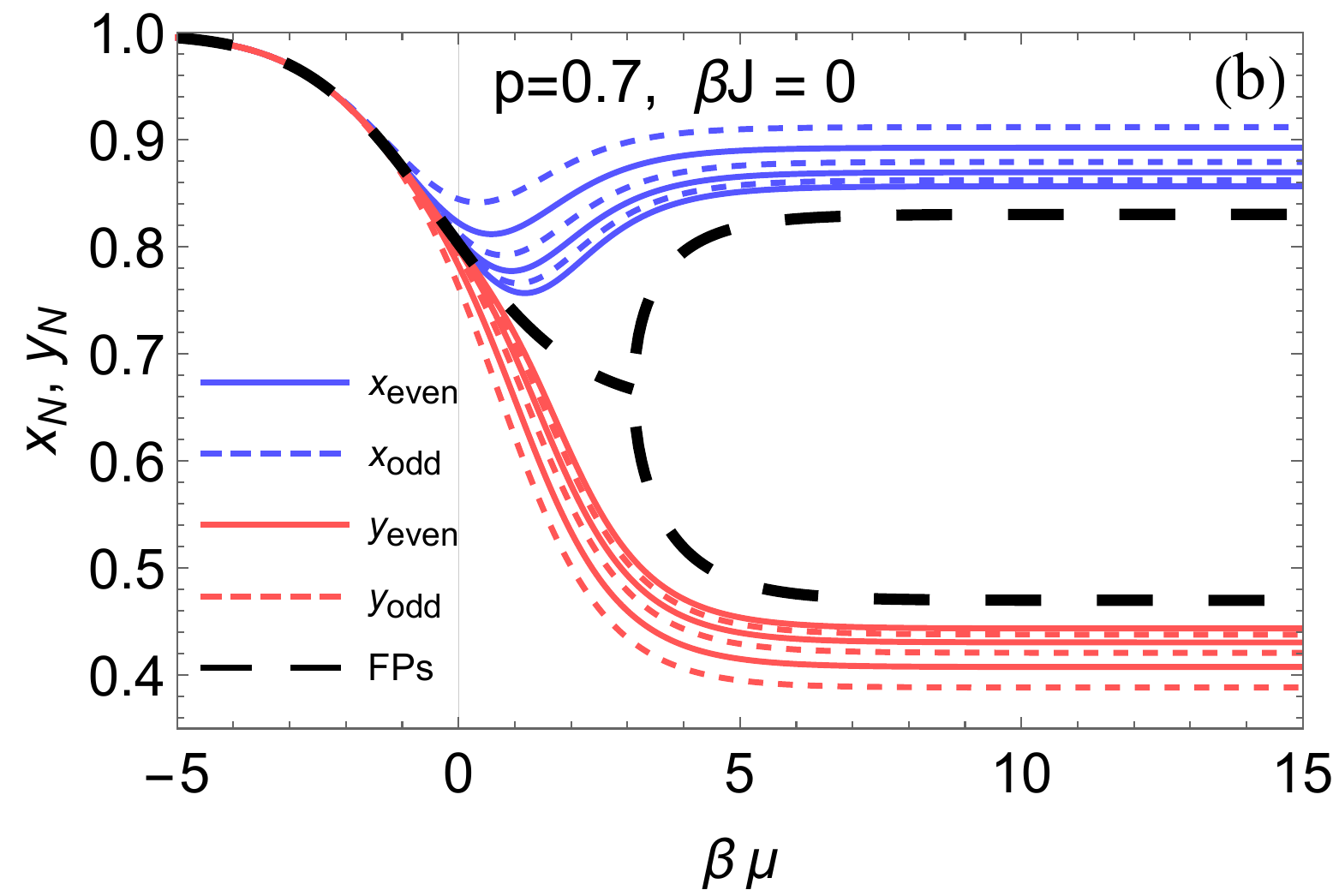} 
		\end{center}
		\caption{(Colour online) Solutions
		$x_{N} = x_{N}(J, \mu, p)$ (blue curves) and  $y_{N} = y_{N}(J, \mu, p)$ (red curves) of 
		the recursion relations \eqref{FinalCayleyRec2} with $J = 0$ and $q = 3$ for $N = 3,  \ldots 8$, plotted as function of $\beta \mu$.  Solid (dashed) curves correspond to even (odd) $N$. The initial values are $x_{0} = 1$ and $y_{0} = 0$.  Thick dashed curves (black) present the limiting  ($N \to \infty$) solutions $x$ and $y$ - the fixed point (FP) solutions. Panel (a): $p = 0.3$. $x_N$ and $y_N$  converge to the same value $x =y$ in the limit $N \to \infty$ (see equation \eqref{dis1a}). Panel (b): $p =0.7$. Spontaneous breaking of a symmetry at $\beta \mu_{c}$, equation \eqref{criticalz0}, followed by a transition into the PBS -- $x_N$ and $y_N$ converge to limiting values $x \neq y$ (see equations \eqref{ord1.1a} and \eqref{ord1.2a}).
		}
		\label{rec-rel1}
	\end{figure}

To illustrate the above general discussion with a particular example, let us consider the case $J = 0$, i.e., the case in which the similar species do not have any other interaction between themselves apart from the hard-core one.  In this case, evidently, one may have only a continuous transition because $(\beta J)_{tc}$ is always greater than zero. Moreover, a continuous transition may only occur if $J(p) < 0$, which imposes some restrictions on the value of $p$: it has to exceed the critical value  $p^{\rm (Bet)} = 2/3$, (where the superscript ${\rm (Bet)}$ signifies that this critical value is specific to the Bethe lattice). Otherwise, for $p < p^{\rm (Bet)}$ the system will not exhibit
 any transition for any value of the chemical potential and will remain in the symmetric phase. For $p > p^{\rm (Bet)}$, conversely, there will occur a continuous transition into the phase with a broken symmetry for $z$ equal to
 \begin{eqnarray}
\label{criticalz0}
z_c = e^{(\beta \mu)_{c}} = \frac{9}{4 (3p -2)}.
\end{eqnarray}
 In Fig.~\ref{rec-rel1} we illustrate different kinds of a  
convergence to the limiting behaviour in the cases when $p <  p^{\rm (Bet)}$ or $p > p^{\rm (Bet)}$. To this end, we present the solutions  $x_N$ and $y_N$ of recursion relations   
 \eqref{FinalCayleyRec2} 
 with $J = 0$ and $q = 3$, for 
 several low order values of $N$,
 $N = 3,  \ldots 8$. We observe that  for $p < p^{\rm (Bet)}$ (Fig.~\ref{rec-rel1}, panel (a)) the solutions $x_N$ and $y_N$ converge with a growth of $N$ to 
 the same ultimate value $x=y$ --  the fixed point solution, 
  depicted by a thick dashed black curve --
  for any value of $\beta \mu$. Moreover, we conclude that 
 only the lowest order solutions deviate from the fixed point solution in a noticeable way;
  $x_1$ (dashed blue curve) and $x_2$ (solid blue curve) appear to be slightly 
  above the latter, while $y_1$ (dashed red curve) and $y_2$ (solid red curve)  are slightly below the fixed point solution. The solutions with larger $N$ get progressively closer and eventually become almost indistinguishable from the fixed point solution.  Concurrently, for $p > p^{\rm (Bet)}$ (Fig.~\ref{rec-rel1}, panel (b)) such a convergence takes place only for moderate values of $\beta \mu$, i.e., for $\beta  \mu < (\beta \mu)_c$, equation \eqref{criticalz0}. For $\beta \mu$ exceeding the critical value in equation \eqref{criticalz0}, a breaking of a symmetry between $x$ and $y$ takes place such that 
  $x_N$ approach from above the upper branch of the fixed point solution (here, the curves for progressively larger $N$ are ordered from top to bottom), while $y_N$ approach from below the lower branch (here, the curves for progressively larger $N$ are ordered from bottom to top).  We note that the fact that, upon a breaking of the symmetry, $x_N$  appears to be larger than $y_N$ (but not vice versa), is due to the initial condition that we have chosen here; that being,  $x_0 = 1$ and $y_0 = 0$. We note, as well, that the convergence in the region with a broken symmetry is visibly much slower. Indeed, the solutions with the largest considered $N$, i.e., $x_{8}$ and $y_{8}$, are still quite far from the fixed point solutions even far away from the transition point. Evidently, the convergence is slowest in the vicinity of the latter. 
 
\begin{figure}%[h!]
	\begin{center}
		\includegraphics[width=0.47\hsize]{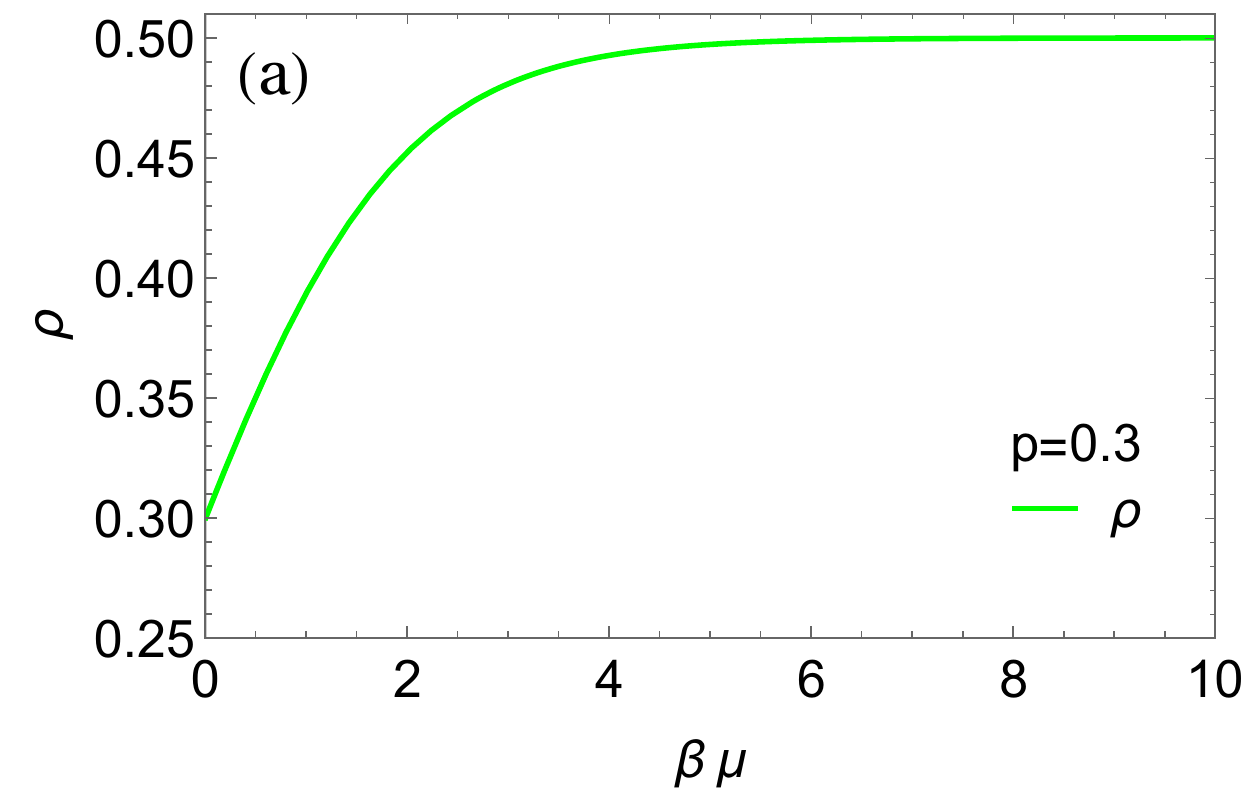} \hfill
		\includegraphics[width=0.47\hsize]{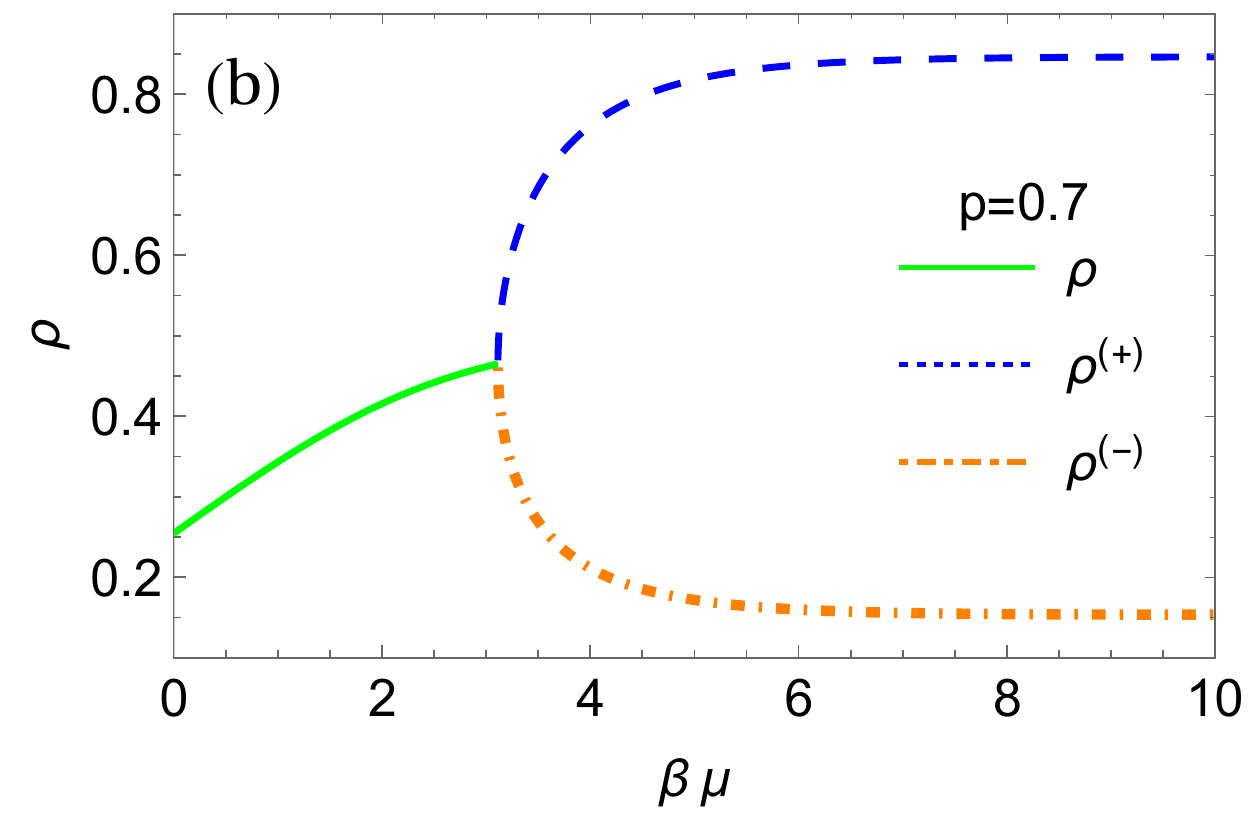}
	\end{center}
	\caption{(Colour online). Mean densities of hard-core ($J = 0$) $A$ and $B$
		particles on the Bethe lattice with $q = 3$ as functions of $\beta \mu$ for two values of  $p$. Panel (a): $p = 0.3 < p^{\rm (Bet)}$  and panel (b): $p = 0.7> p^{\rm (Bet)}$. The solid (green) curve shows the mean particles' density  $\rho = \rho^{(A)} = \rho^{(B)} $ in the symmetric phase, in which they are equal to each other.  The dashed (blue) and the dash-dotted (orange) curves
		in panel (b) show the mean particles' densities $\rho^{(+)}$ and $\rho^{(-)}$ within the PBS (see equations \eqref{pm}).}
	\label{fig3}
\end{figure} 
  
 Such a behaviour of solutions of the recursion relations  \eqref{FinalCayleyRec2}, in virtue of equation \eqref{meandens4} relating $x$ and $y$ to the mean densities, is evidently translated 
  into a similar behaviour of the latter. 
In Fig.~\ref{fig3} we depict the dependence of the mean densities on $\beta \mu$ for these chosen cases. We observe that for $p = 0.3$, (which is below $p^{\rm (Bet)} = 2/3$), the densities of both species are equal to each other and  increase monotonically from zero to the limiting value $1/2$, (when $\beta \mu$ is varied from $-\infty$ to $+ \infty$), showing that the repulsive interactions between the dissimilar species are not sufficiently strong to prevent a complete coverage of the system. The penalty one has to pay for having an $A$ and a $B$ at the neighbouring sites is paid here by the chemical potential. Conversely, for $p = 0.7$, (which exceeds $p^{\rm (Bet)} = 2/3$), the situation appears to be different; here, the mean densities of both species are equal to each other
for moderate values of $\beta \mu$,  and then, when $\beta \mu$ approaches a critical value $(\beta \mu)_c$ (see equation \eqref{criticalz0}), a spontaneous symmetry-breaking occurs and within the PBS the densities are no longer equal. If we keep on increasing $\beta \mu$, both densities will approach their ultimate values that depend on the value of $p$:
\begin{eqnarray}
\label{pm}
	\rho^{(+)} \Big|_{(\beta \mu) \to \infty} &=& \frac{1}{2}+\frac{1}{2}\frac{\sqrt{(8-3 p) p-4}}{\Big((5-2 p) p-2\Big)}, \nonumber\\
	\rho^{(-)} \Big|_{(\beta \mu) \to \infty} &=& \frac{1}{2}-\frac{1}{2}\frac{\sqrt{(8-3 p) p-4}}{\Big((5-2 p) p-2\Big)}, 
\end{eqnarray}
where the function under the radical
is positive for $p > p^{\rm(Bet)}$ and vanishes at $p = p^{\rm (Bet)}$. It is worthy to note that $\rho^{(+)}$ ($\rho^{(-)}$) is a monotonically increasing (decreasing) function of $p$ and the maximal value $\rho^{(+)} = 1$ ($\rho^{(-)} = 0$) is achieved only for $p=1$ and $\beta \mu = \infty$, i.e. for an infinitely strong repulsion between dissimilar species and for an infinitely large chemical potential. However, for $p^{\rm (Bet)} < p < 1$, the breaking of the symmetry is not complete even for $\beta \mu \to \infty$ and some amount of the minority species is present. We also note parenthetically that  values of the mean densities corresponding to zero chemical potential (i.e., for $z = 0$) only very slightly deviate from 
$1/3$ for both cases: $\rho \approx 0.2993$ (for $p=0.3$) and $\rho \approx 0.2543$ (for  $p = 0.7$) (see also \ref{AppC}).

\begin{figure}%[h]
	\begin{center}
		\includegraphics[width=0.5\columnwidth]{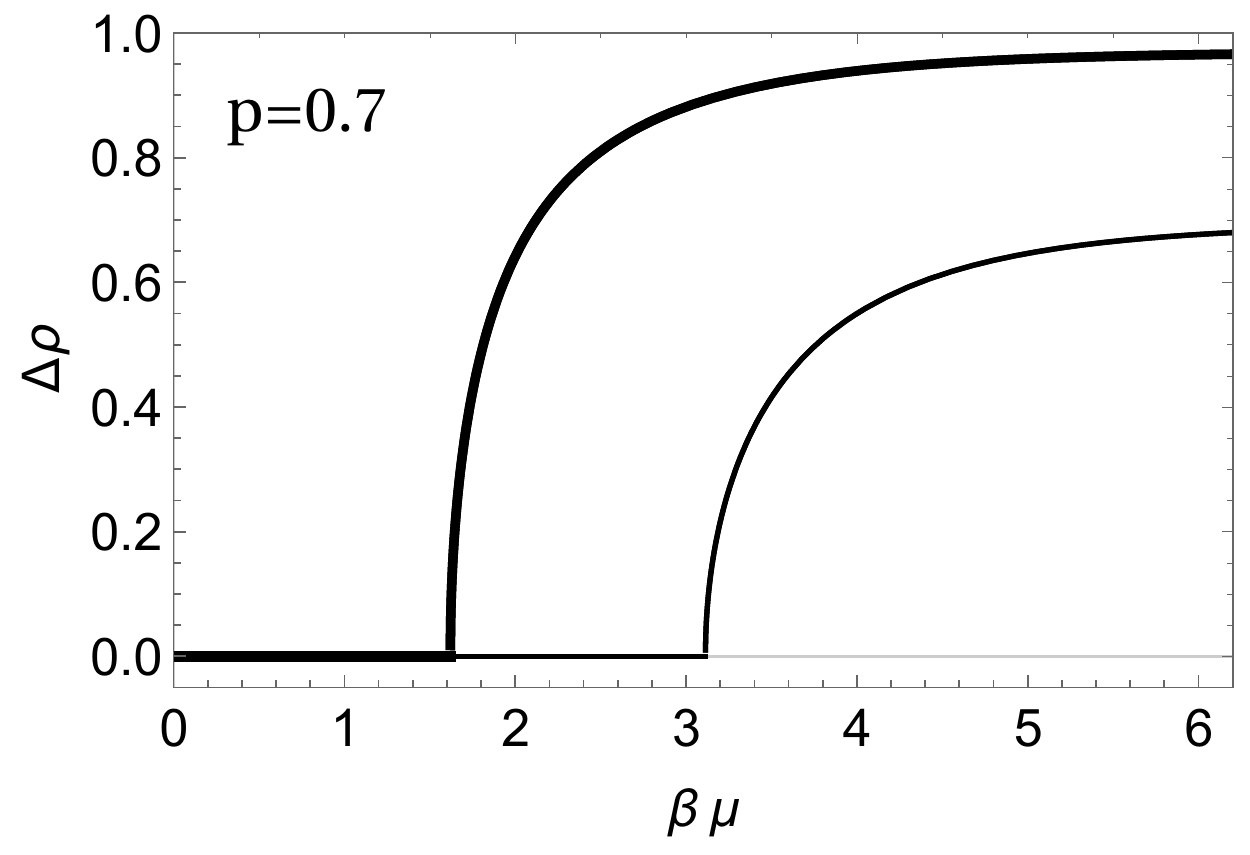}
	\end{center}
	\caption{Order parameter $ \Delta \rho $ for $J = 0$ and $p = 0.7$ as function of $ \beta \mu$. Thin
	curve shows the result for the Bethe lattice (see equation \eqref{OP} in the \ref{A}), while the thick curve presents an analogous result for the Husimi lattice, equation \eqref{orderhus}.}
	\label{fig4}
\end{figure}

Further on, it might  also be instructive to check how fast the densities depart from each other upon a transition into the PBS. To this end, we
depict in Fig. \ref{fig4} (thin curve)
the absolute value $\Delta\rho$ of the difference of $\rho^{(A)}$ and $\rho^{(B)}$, $\Delta\rho = |\rho^{(A)} - \rho^{(B)}|$, as a function of $\beta \mu$, which is the natural order parameter in the model under study.
Obviously, $\Delta\rho$, which is explicitly defined in equation \eqref{OP} in  \ref{A}, is identically equal to zero in the symmetric phase and becomes non-zero in the PBS. We observe that the growth of $ \Delta \rho $ for $\mu > \mu_c$ is rather steep; we show analytically
in \ref{scaling}
that the order parameter behaves as $\Delta \rho \sim (\mu - \mu_c)^{1/2}$ in the vicinity of $\mu = \mu_c$. As we have already remarked, this is a mean-field-type prediction for the value of the critical exponent in case of a continuous transition, which is the consequence of the fact that the Bethe-lattice is an effectively infinitely dimensional system. In \ref{scaling} we show, as well, that when a transition into the PBS takes place at the line of the tricritical points, $\Delta \rho \sim (\mu - \mu_c)^{1/4}$ in the vicinity of $\mu = \mu_c$, which is another well-known 
"mean-field "value of the critical exponent.

\subsection{Symmetric phase with a structural order}

\begin{figure}%[htbp]
		\begin{center}
			\includegraphics[width=0.47\hsize]{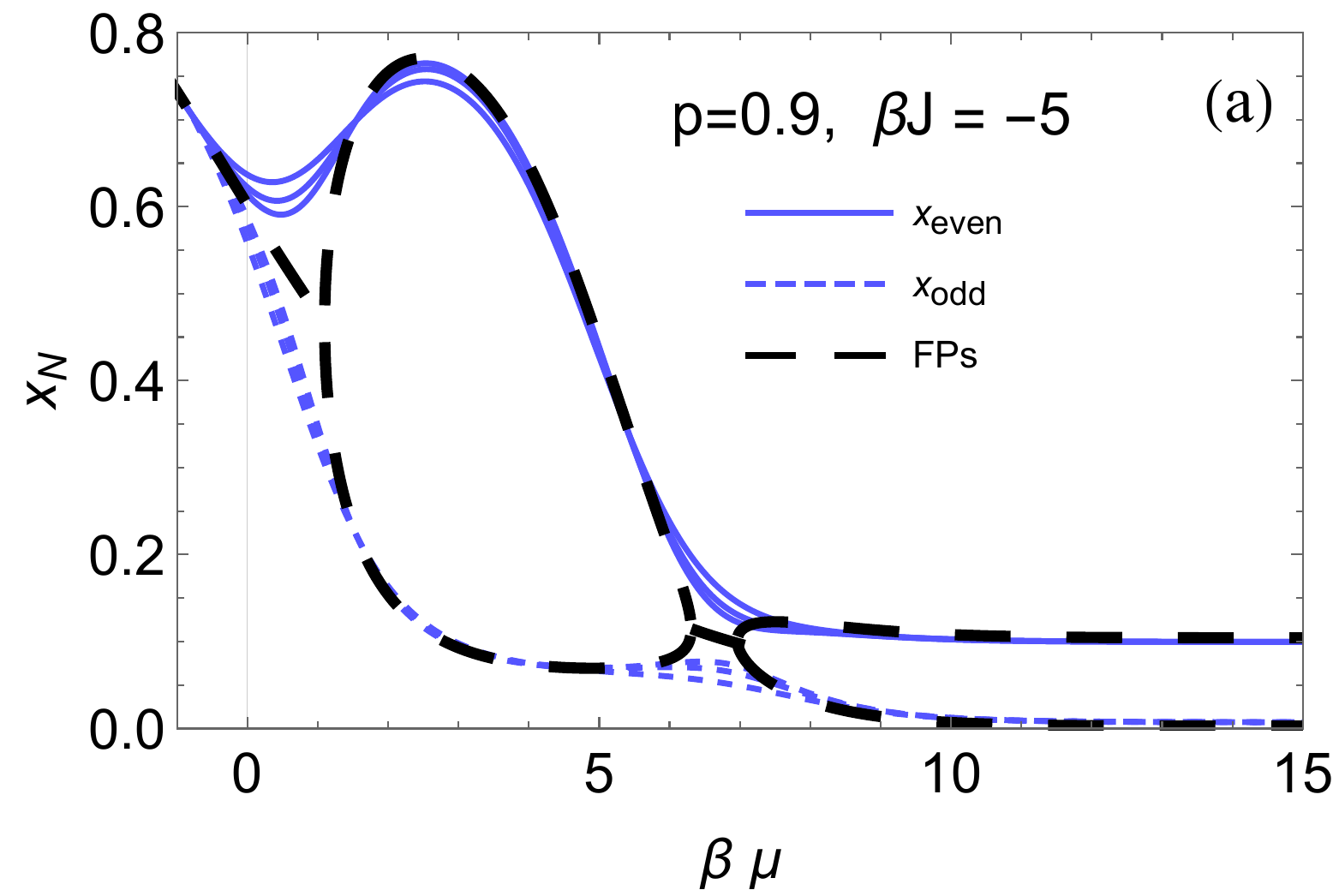}  \hfill
			\includegraphics[width=0.47\hsize]{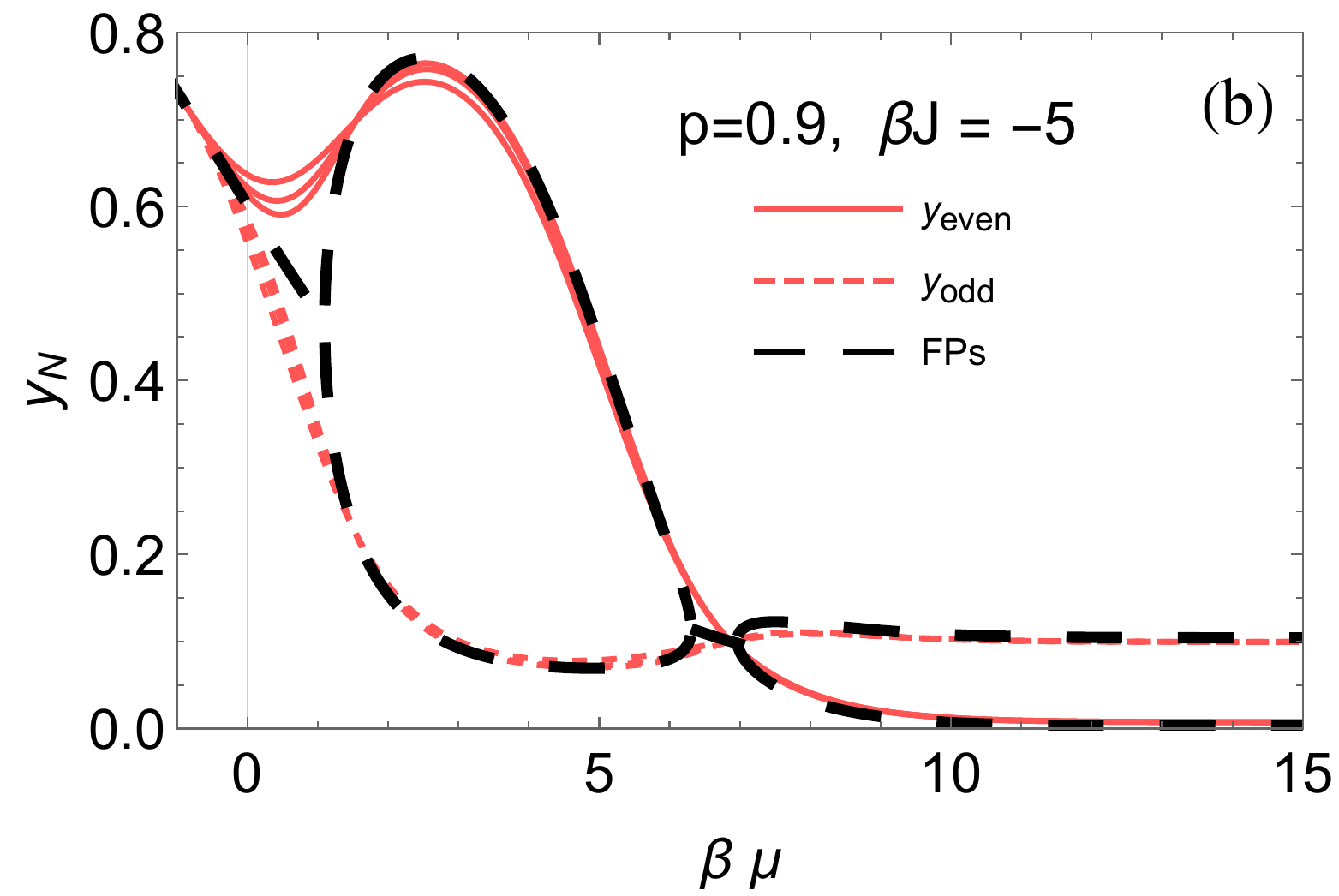}
		\end{center}
		\caption{(Colour online) Solutions $x_{N} = x_{N}(J, \mu, p)$ (panel (a)) and $y_{N} = y_{N}(J, \mu, p)$  (panel (b)) of the recursion relations \eqref{FinalCayleyRec2} with $q = 3$, $p=0.9$, and $\beta J = -5$, plotted as functions of $\beta \mu$. Initial conditions are $x_{0} = 1$ and $y_{0} = 0$.  Solid curves correspond to even $N$, the dashed ones - to odd $N$, respectively, while the thick (black) dashed curve depicts the fixed point solution. For $\beta \mu < (\beta \mu)_{c,1}(p,J)$ and $(\beta \mu)_{c, 2} (p, J) < \beta \mu < (\beta \mu)_{c, 3} (p, J)$ (see equations \eqref{zc1}, \eqref{zc2} and \eqref{zc3}), $x_N$ and $y_N$ converge to the same fixed point solution $x = y$, regardless of the parity of $N$, which corresponds to the symmetric disordered phase. Within the interval  $(\beta \mu)_{c, 1} (p, J) < \beta \mu < (\beta \mu)_{c, 2} (p, J)$, $x_N$ and $y_N$ with odd $N$ converge to  $x_{odd} = y_{odd}$ as $N \to \infty$, while the solutions with even $N$ converge in this limit to $x_{even} = y_{even}$, respectively. This region defines the alternating order phase PAO I, in which $A$s and $B$s occupy predominantly one of the sub-lattices, leaving the second one almost empty.
 For $ \beta \mu > (\beta \mu)_{c, 3} (p, J)$ a different kind of a convergence takes place when $x_N$ with \textit{even} $N$ and $y_N$ with \text{odd} $N$ converge to the same limiting curve $x_{even} = y_{odd}$,  while $x_N$ with \textit{odd} $N$ and $y_N$ with \text{even} $N$ converge to 
 $x_{odd} = y_{even}$. This region defines 
 the alternating order phase -- PAO II, in which 	similar species
 occupy predominantly the same sub-lattice. }
		\label{rec-rel2}
	\end{figure}

We note now that the discussion in the previous subsection does not provide
an exhaustive picture, as one may infer, e.g., from  Fig. \ref{rec-rel2}, in which we depict low order ($N =3,  \ldots, 8$) solutions of recursion relations
 \eqref{FinalCayleyRec2} with $q = 3$, $p=0.9$ and $\beta J = -5$, (i.e., there is a rather strong repulsion between both similar and dissimilar species), plotted as functions of $\beta \mu$.  
Inspecting the recursion relations \eqref{FinalCayleyRec2} further, we realise that  there exist two regions in the parameter space situated well within the symmetric phase, (both regions emerge at sufficiently large negative $\beta J$), in which $x_N$ and $y_N$ with $N$ odd and with $N$ even converge to different limiting curves, while we still have $x_N = y_N$ as $N \to \infty$, i.e., the mean densities of both species are the same. 
In other words, in these regions 
 there is no breaking of a symmetry between the mean particles' densities as observed in Sec. \ref{qq}, but instead some kind of a structural order emerges, that manifests itself via a breaking of the symmetry between the solutions that have a different parity.  As we have already mentioned, the point is that here the system (recall that the Bethe lattice is a bipartite lattice) partitions spontaneously 
into two different sub-lattices with different particles' arrangements on each of them, and solutions with even $N$ define the occupation of one of the sub-lattices, while the solutions with odd $N$ - the occupation of the other one. 
We show in what follows that these special regions correspond  to the phases with an alternating order: one of them is the phase  (PAO I) in which both $A$s and $B$s appear predominantly on the same sub-lattice, 
 leaving the second sub-lattice almost empty,  while in the second phase (PAO II) the $A$ 
 particles occupy predominantly one sub-lattice, while $B$s appear mostly on the second one.
 
 More specifically, 
there exist (see \ref{A} for the derivation) three critical values of the activity $z$ (the logarithms of which define the corresponding critical values of the chemical potential),  
\begin{flalign}
\label{zc1}
z_{c, 1}(p, J) {=} \frac{3 p (6 {-} p) {-} 11 {-} 3 \, e^{2 \beta J}  {-} 6 (3 {-} p) e^{\beta J} {-} \sqrt{\left(9 p {-} 7 {-} 9 \, e^{\beta J}\right) \left(1 {+} p {-} e^{\beta J}\right)^3}}{8 \left(1 {-} p {+} e^{\beta J} \right)^3}, \\
\label{zc2}
z_{c, 2}(p, J) {=} \frac{3 p (6 {-} p) {-} 11 {-} 3 \, e^{2 \beta J}  {-} 6 (3 {-} p) e^{\beta J} {+} \sqrt{\left(9 p {-} 7 {-} 9 \, e^{\beta J}\right) \left(1 {+} p {-} e^{\beta J}\right)^3}}{8 \left(1 {-} p {+} e^{\beta J} \right)^3},
\end{flalign}
and
\begin{flalign}
		\label{zc3}
		z_{c, 3}(p,J) = \frac{(3 e^{\beta  J} + p - 3)^{2}}{4 (e^{\beta  J} + p - 1)^{2} (1 - 3 e^{\beta  J} - p)},
\end{flalign}
which delineate the boundaries of the phases with an alternating order. 
We now dwell some more on the loci and properties of these phases.

(i) The disordered symmetric phase exists only when $\beta \mu < (\beta \mu)_{c,1}(p,J)$ or when $\beta \mu$ is within the bounded interval $(\beta \mu)_{c, 2} (p, J) < \beta \mu < (\beta \mu)_{c, 3} (p, J)$. In this range of values of the chemical potential solutions of the recursion relations \eqref{FinalCayleyRec2},
regardless of the parity of $N$,  converge to the same value $x = y$ which is the fixed point solution. This implies that here, in virtue of equation \eqref{meandens4}, $\rho^{(A)} = \rho^{(B)}$.

(ii) When $\beta \mu$ exceeds $(\beta \mu)_{c,1}(p,J)$ the system enters, via a continuous transition, into the PAO I in which the symmetry between solutions with a different parity is broken. Indeed, here we have that $x_N$ ($y_N$) with even $N$ and with $N$ odd converge to different limiting curves -- $x_{even}$  ($y_{even} = x_{even}$) and $x_{odd}$  ($y_{odd} = x_{odd}$). For example, we observe in Fig.  \ref{rec-rel2} that for $\beta \mu \approx 3$ the solutions
$x_{even}$ and $y_{even}$ attain their maximal value $x_{even} = y_{even} \approx 0.8$, while $x_{odd}$ and $y_{odd}$ are close to their minimal value,  $x_{odd} = y_{odd} \approx 0.1$. This implies that, in virtue of equation \eqref{meandens4}, 
on one of the sub-lattices we have $\rho^{(A)} = \rho^{(B)} \approx 0.48$, i.e., a very high coverage by both species which are present in equal amounts, 
while on the second one we have $\rho^{(A)} = \rho^{(B)} \approx 0.02$, i.e., this sub-lattice is almost completely devoid of particles. 
The system leaves this phase and re-enters the symmetric disordered phase, again via a continuous transition, when $\beta \mu =  (\beta \mu)_{c,2}(p,J)$. 

We note now that $z_{c, 1}(p, J)$ and  $z_{c, 2}(p, J)$ in equations
\eqref{zc1} and \eqref{zc2}
have to be real positive numbers. This latter condition implies that there are
some restrictions on the values of $p$ and $J$. Namely, the PAO I may only exist
when $p$ and $\beta J$ obey simultaneously 
\begin{eqnarray}
p \geq p_c = \frac{7}{9} + e^{\beta J} \,, \quad 
\beta J \leq \ln\left(2/9\right) \,, 
\end{eqnarray}
which inequalities define the location of this  phase on the phase diagram in Fig. \ref{ph-dim}.

(iii) At $\beta \mu = (\beta \mu)_{c,3}(p,J)$ the system enters, via a continuous transition, from the symmetric disordered phase into the PAO II and stays within this phase for any $\beta \mu \in ((\beta \mu)_{c,3}(p,J),\infty)$, i.e., extends to infinitely large values of the chemical potential. As can be seen in Fig. \ref{rec-rel2}, 
in this phase a salient feature is that, while the solutions $x_N$ with $N$ even and with $N$ odd maintain their order in the sense that $x_{even} > x_{odd}$, likewise it happens within the PAO I, the solution $y_N$ with $N$ even chooses here the lower branch of the fixed point solution, while the solution with $N$ odd selects the upper branch, such that  $y_{odd} > y_{even}$ in the PAO II. Hence, there is a spontaneous breaking of the symmetry between the solutions with a different parity such that the system partitions into two sub-lattices, but particles' arrangements on each of the sub-lattices is completely different,  as compared to the one in PAO I.  Suppose that we take $\beta \mu = 10$, which is well within this phase. Then,  we have that here $x_{even} = y_{odd} \approx 0.09$ and $x_{odd} = y_{even} \approx 0.01$. From our equation \eqref{meandens4} it follows  then that on one of the sub-lattices $\rho^{(A)} \approx 0.48$ and $\rho^{(B)}  \approx 0.02$, while on the other - $\rho^{(A)} \approx 0.02$ and $\rho^{(B)}  \approx 0.48$. Therefore, the PAO II is the phase with an alternating structural order but here the system splits into two sub-lattices  each of which is predominantly occupied by just one kind of species.  

For the PAO II to exist, it is necessary that the critical activity $z_{c, 3}(p,J)$ in equation \eqref{zc3} is a real positive number.  This can only be realised for such values of $p$ and $J$ which obey the inequality 
\begin{align}
1 - 3 e^{\beta  J} - p \geq 0 \,,
\end{align}
which, together with the condition $\beta \mu \geq (\beta \mu)_{c,3}(p,J)$, defines the location of the PAO II on the phase diagram in  Fig.~\ref{ph-dim}. 

Lastly, we note that the passage
from the PAO I to PAO II, upon an increase of the chemical potential, proceeds via a transition through the symmetric disordered phase. Leaving the PAO I, the system thus looses the structural order I and becomes disordered. It regains a structural order of a different kind upon entering the PAO II.  One can straightforwardly verify that the difference 
$(\beta \mu)_{c,3}(p,J) - (\beta \mu)_{c,2}(p,J)$, which defines the range of values of the chemical potential 
in which the system is in the disordered phase, is always positive and finite. This difference vanishes only 
for systems with $p = 1$, (i.e., for an infinitely strong repulsion between dissimilar species), when, additionally, one goes to the limit of an infinitely strong repulsion between similar species, i.e., $J \to - \infty$.

%%%%%%%%%%%%%%%%%%%%%%%%%%%%%%%%%%%%%%%%%%%%%%%%%%%%%%%%%%%%%%%%
\section{The Husimi lattice}
\label{husimi}
%%%%%%%%%%%%%%%%%%%%%%%%%%%%%%%%%%%%%%%%%%%%%%%%%%%%%%%%%%%%%%%%

In this section we consider our model on the Husimi tree (see Fig. \ref{fig1}, panel (b)). In our analysis, we proceed along the same lines as it was done in case of the Bethe lattice. We first evaluate appropriate recursion relations, obeyed by the partition function and then turn to the limit $N \to \infty$ concentrating on the behaviour of the interior sites which are far away from the boundary. This geometrical construction represents the so-called Husimi lattice. Our derivations of the recursion relations are performed for arbitrary $J_A$, $J_B$, $z_A$ and $z_B$, and $t$ - the number of triangles that meet each other at each vertex.  The final results are presented and discussed solely for the symmetric case with $J_A = J_B = J$ and $z_A=z_B = z$, and also for the simplest non-trivial geometry with $t = 2$, which corresponds to an approximation of the so-called kagome lattice.

We first write formally $Z(p)$ in equation~\eqref{averagedZnCayley} defined on the Husimi tree in form of equation~\eqref{ZCayley}, i.e. consider three possible events with respect to the occupation of the root site,
\begin{equation}
Z(p) = D_{N}^{t}(0,p) + z_{A}^{1-t} D_{N}^{t}(A,p) + z_{B}^{1-t} D_{N}^{t}(B,p) \,,
\end{equation}
where $D_{N}(0,p)$, $D_{N}(A,p)$ and $D_{N}(B,p)$ are, respectively, the
grand canonical partition functions of a single branch with a root site which is vacant, occupied by an $A$ particle, or occupied by a $B$ particle. Recursion relations obeyed by these conditional grand canonical partition functions are listed in the \ref{B} (see equations \eqref{d0} and \eqref{dA}).

Introducing next auxiliary variables
\begin{eqnarray}
\label{newvar_h}
x_{N} = \frac{D_{N}(A,p)}{z_{A}D_{N}(0,p)}, \qquad {\text{and}} \qquad y_{N} = \frac{D_{N}(B,p)}{z_{B}D_{N}(0,p)},
\end{eqnarray}
we find that they obey (for arbitrary $J_A$, $J_B$, $z_A$, $z_B$ and $t$) the following recursions:
\begin{eqnarray}
\label{systeqH}
 {x}_{N} &{=}&\frac{ 1 {+} 2 e^{\beta J_A} \xi  {+}2  (1 {-} p) \eta  {+}2  (1 {-} p)^{2} e^{\beta J_A}  \xi\eta{+}   e^{3 \beta J_A}  \xi^2 {+} (1{-}p)^{2} e^{\beta J_B}\eta^2  }{  1 {+} 2 \xi  {+}2 \eta  {+}2   (1 {-} p) \xi\eta {+}  e^{\beta J_A}  \xi^2  {+}  e^{\beta J_B} \eta^2 }, \nonumber \\
{ {y}}_{N} &{=}& \frac{1 {+} 2(1 {-} p) \xi  {+}2  e^{\beta J_B} \eta  {+}2 (1 {-} p)^{2} e^{\beta J_B} \xi\eta{+}  e^{3 \beta J_B}\eta^2  {+} (1{-}p)^{2} e^{\beta J_A} \xi^2 }{  1 {+} 2 \xi  {+}2 \eta  {+}2   (1 {-} p) \xi\eta {+}  e^{\beta J_A}  \xi^2  {+}  e^{\beta J_B} \eta^2 },
\end{eqnarray}
with notations  $\xi=z_A x_{N-1}^{(t-1)}$ and $\eta=z_B y_{N-1}^{(t-1)}$.  Equations \eqref{systeqH}  have a substantially more complicated form (even in the symmetric case) than their counterparts in equations \eqref{FinalCayleyRec2}, evaluated for the Bethe lattice.

We next turn to the limit $N \to \infty$ and consider only the sites which are deep inside the Husimi tree, (i.e. belong to the Husimi lattice). Then, we realise that the recursion relations \eqref{systeqH} converge to some fixed point solutions $x$ and $y$, which may be equal to each other or have unequal values, and thus correspond to different thermodynamic phases. From now on we concentrate on the symmetric case and also set $t =2$.

As in case of the Bethe lattice (see the \ref{A}), the subsequent analysis is conveniently performed in terms of variables $u = (x + y)/2$ and $v = (x - y)/2$. Changing the variables $x$ and $y$ in equations \eqref{systeqH} for $u$ and $v$, we find that the latter obey  non-linear equations of the form
\begin{align}
\label{symm_sysJ}
u  &{=} \frac{1 {+} 2 z \left(1{-}p {+} e^{\beta J}\right) u  {+} 2 z^{2} e^{\beta J} \left(e^{2 \beta J} {+} (1{-}p)^{2}\right) u ^{2} {+} z^{2} e^{\beta J} \left((1{-}p)^{2} {-} e^{2 \beta J}\right)\left(u ^{2} {-}  v^{2}\right)}{1 {+} 4 z u  \left(1  {+} z e^{\beta J} u  \right) {+} 2 z^{2} \left((1 {-} p) {-} e^{\beta J}\right) \left(u^{2} {-}  v^{2}\right)}, &&\nonumber \\
 v &{=} 2 z \left(e^{\beta J} {-} (1 {-} p)\right)  v \frac{1 {+} z e^{\beta J} \left(1 {-} p {+} e^{\beta J}\right) u}{1 {+} 4 z u \left(1  {+} z e^{\beta J} u \right) {+} 2 z^{2} \left((1 {-} p) {-} e^{\beta J}\right) \left(u^{2} {-}  v^{2}\right)}.&&
\end{align}
The system of equations \eqref{symm_sysJ} has two following solutions: \\
(i) the solution with $v = 0$ is defined by
\begin{eqnarray}
\label{disH}
 2 z^{2} \left(1 - p + e^{\beta J} \right) u^{3} &+& z \left(4 - z e^{\beta J} \left(e^{2 \beta J} + 3 (1-p)^{2}\right)\right) u^{2} \nonumber \\ &+& \left(1 - 2 z \left(1 - p + e^{\beta J} \right) \right) u - 1 = 0.
\end{eqnarray}
This solution evidently corresponds to the symmetric phase in which the $A$ and $B$ particles are present at equal mean densities. An analogous phase observed on the Bethe lattice was coined a disordered symmetric one;\\
(ii) the solution with $ v \neq 0$ is defined by a pair of equations
\begin{eqnarray}
\label{ordH}
&&2 z^{2} e^{\beta J} \left(2 (1-p) + e^{\beta J}\right) \left(e^{\beta J} - (1-p)\right) u^{2} \nonumber \\ &-& z\left(4 (1-p) - e^{\beta J} \left(e^{\beta J} + 1 - p\right)\left(2 - z e^{\beta J} \left(e^{2\beta J} - (1-p)^{2}\right)\right)\right) u\nonumber \\ &+& \frac{1}{2} e^{\beta J} \left(e^{\beta J} + 1-p\right)\left(1 - 2 z \left(e^{\beta J} - (1-p)\right)\right) - 1 = 0,
\end{eqnarray}
and
\begin{flalign}
\label{phase2J_2}
\frac{1}{2 z} {+} 2 u (1 {+} z e^{\beta J} u) {+} z \left(1{-}p {-} e^{\beta J}\right) \left(u^{2} {-} v^{2}\right) {=} \left(e^{\beta J} {-} (1-p)\right) \left(1 {+} z e^{\beta J} \left(1 {-} p {+} e^{\beta J}\right)u\right).
\end{flalign}
This solution corresponds to the phase with a broken symmetry, in which the mean densities of the species are no longer equal to each other.

Further on, a parametric equation which defines implicitly  the location of a part (corresponding to a continuous transition) of the
surface separating these two phases,
can be obtained by assuming that equations~\eqref{disH} and~\eqref{ordH} are  fulfilled simultaneously.  This yields a rather lengthy expression~\eqref{critical_eqJ}, which is presented in the \ref{B}, together with the corresponding exact expression for the critical value of the activity, equation~\eqref{critical_activity__huisimi}. In turn, the  part of such a surface corresponding to the first order transition is obtained in a standard way by equating the free energies (see equation~\eqref{FreeEnergyH}) of the symmetric phase and of the phase with a broken symmetry. This finally yields equation~\eqref{tricritical_eq__husimi}, which defines the line of the tricritical points. Naturally, in view of a more complicated form of the recursion relations obeyed by the partition function of the Husimi tree, the result in equation \eqref{tricritical_eq__husimi} is much more complicated than its counterpart in equation~\eqref{jtr} which is valid for the Bethe lattice.  
 On the contrary, a critical parameter $\beta J(p)$ in case of the Husimi lattice is simply given by
\begin{eqnarray}
\label{Jph}
	\beta J(p) = \ln{(\sqrt{5}(1-p))} \,,
\end{eqnarray}
and hence, it differs from its counterpart for the Bethe lattice in equation~\eqref{Jp} only by a numerical factor. Recall that no transition takes place for $J {<} J(p)$ such that the system is always in the disordered symmetric phase.

\begin{figure}%[htbp]
	\begin{center}
		\includegraphics[width=0.65\hsize]{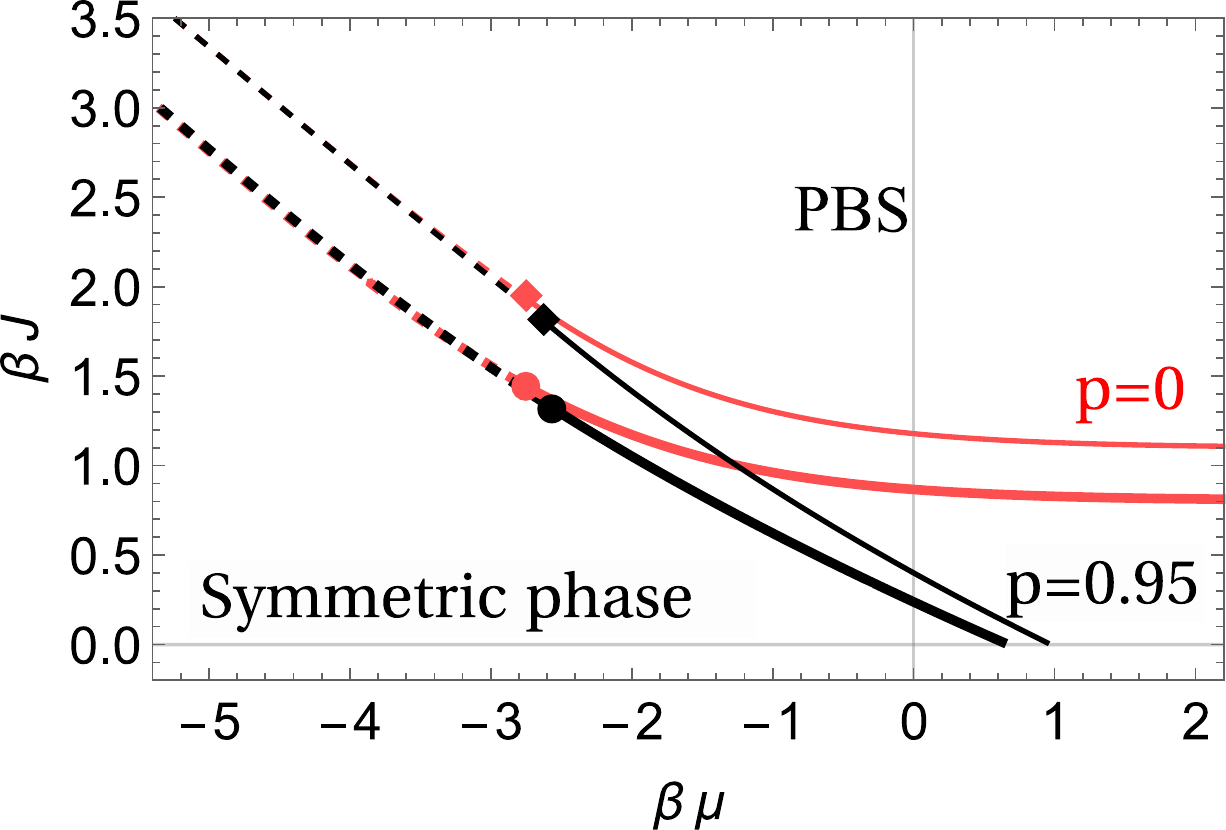}
	\end{center}
	\caption{(Colour online) Phase diagram of the model on the Husimi lattice (thick curves) and on the Bethe lattice  (thin curves) for two fixed values of the parameter $p$. Red curves correspond to $p=0$ (no repulsion between dissimilar species), while black curves - to $p = 0.95$ (strong repulsion between dissimilar species). 
 The solid curves are the lines of a continuous phase transition, while the dashed curves are the lines of a first order phase transition. Note that the
	lines for the first order phase transition are only slightly dependent on the actual value of $p$ in both cases. Lines of first order and second order phase transitions meet at tricritical points denoted by diamonds for Bethe lattice and circles for Husimi lattice. 	}
	\label{phdh}
\end{figure}

In Fig.~\ref{phdh} we depict a phase diagram of the model on the Husimi tree (for comparison, we present it together with its counterpart obtained for the Bethe lattice, which permits us to draw some general conclusions) for two particular values ($p = 0$ and $p=0.95$) of the mean concentration of the catalytic bonds.  We observe that, in general, a transition into the phase with a broken symmetry for systems with the same $J$, (which sets the strength of interactions between similar species), and the same value of $p$, (which defines the strength of repulsive interactions between dissimilar species), occurs on the Husimi tree at lower values of the chemical potential than on the Bethe lattice. Such a behaviour may be apparently attributed to the fact that on the Husimi tree, due to its specific geometry, the system is more frustrated than on the Bethe lattice, such that
it appears somewhat easier to break the symmetry between the species.
 Hence, not counter-intuitively,
the onset of the critical behaviour is shifted towards smaller values of $\beta \mu$.   Further on, we realise that the strength of  repulsive interactions between dissimilar species, does not affect in any noticeable way the location of the line of critical points corresponding to the first order phase transition, both for the Husimi tree and for the Bethe lattice; indeed, we see  that on the Husimi tree for $p=0$, when such repulsive interactions are completely absent, and for $p=0.95$, when such interactions are strong, the thick dashed curves  in Fig. \ref{phdh} almost overlap. The same happens in case of the Bethe lattice (see thin dashed curves in Fig. \ref{phdh}).
On the contrary,  the precise location of the line of critical points corresponding to the continuous transition is very much dependent on the value of $p$, for both the Husimi and the Bethe lattice.
As a consequence, also the loci of the tricritical points depend on $p$.

It might be instructive to consider the phase diagram in more detail for a particular case. To this end, we again concentrate on the limit when $J = 0$.
Similarly to the behaviour on the Bethe lattice, here one has only a continuous transition between a symmetric disordered phase and a phase with a broken symmetry. The critical value $z_c$ of the activity at which such a transition takes place obeys the quadratic equation
\begin{flalign}
\label{critical_eq}
  z_c^{2}  {+} \frac{4 (4 {-} 3 p)(4 {-} 5 p)}{8 (1 {-} p) (4 {-} 5 p (2 {-} p))} z_c + \frac{32 {-} 25 p}{8 (1 {-} p) (4 {-} 5 p (2 {-} p))} {=} 0,
\end{flalign}
whose only real and positive root is given  by:
\begin{equation}
\label{crit_eq}
z_{c}(p) =\dfrac{5\sqrt{p(2-p)}+15p+17}{20\left(p- \dfrac{\sqrt{5}-1}{\sqrt{5}}\right)\left(\dfrac{\sqrt{5}+1}{\sqrt{5}}-p\right)} \,.
\end{equation}
As one can infer from equation~\eqref{crit_eq}, the critical value $z_{c}(p)$ of the activity
  is negative and therefore unphysical for $p$ below the critical value $p^{\rm (Hus)} = (\sqrt{5}-1)/\sqrt{5} \approx  0.553$.
Therefore,  for  $p < p^{\rm (Hus)}$  the system does not undergo any phase transition. Note, that  the phase with a broken symmetry may thus emerge on the Husimi lattice at smaller values of $p$ than in case of the Bethe lattice, because $p^{\rm (Hus)}<p^{\rm (Bet)}$.  

\begin{figure}%[h!]
	\begin{center}
		\includegraphics[width=0.47\hsize]{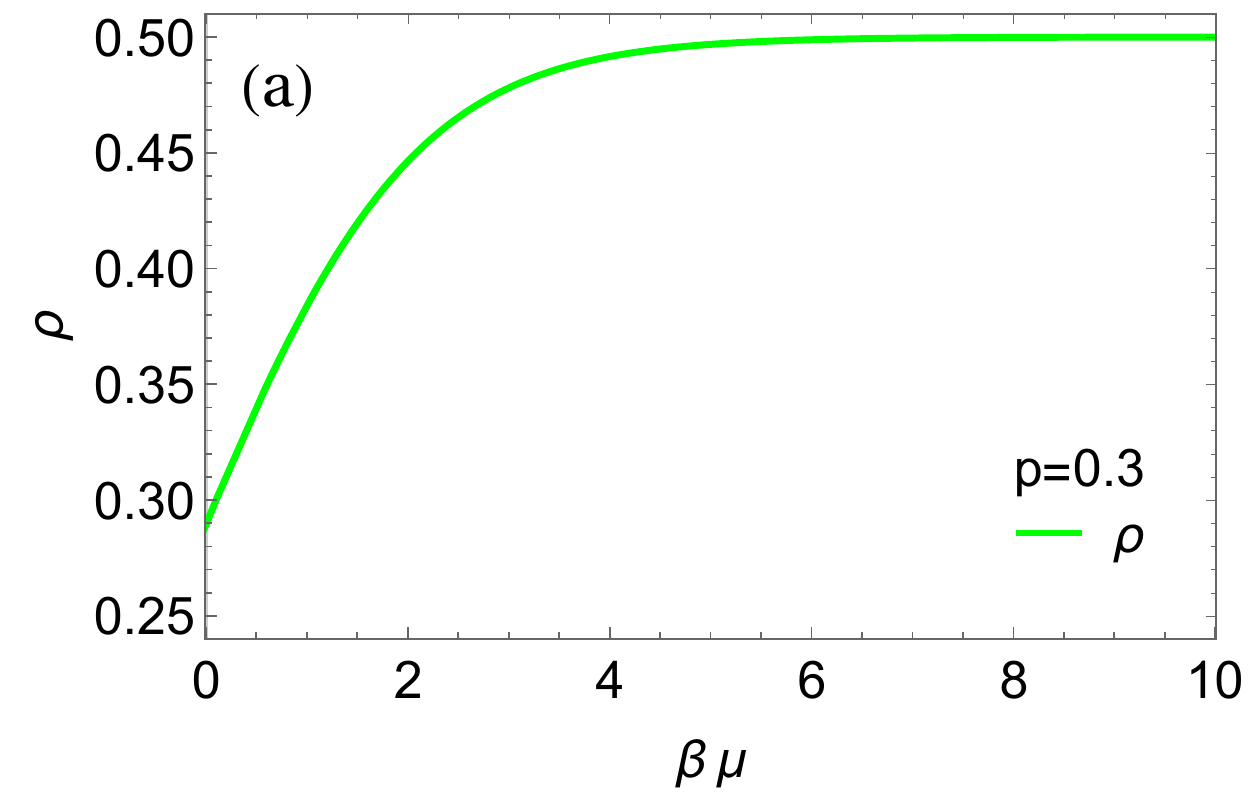} \hfill
		\includegraphics[width=0.47\hsize]{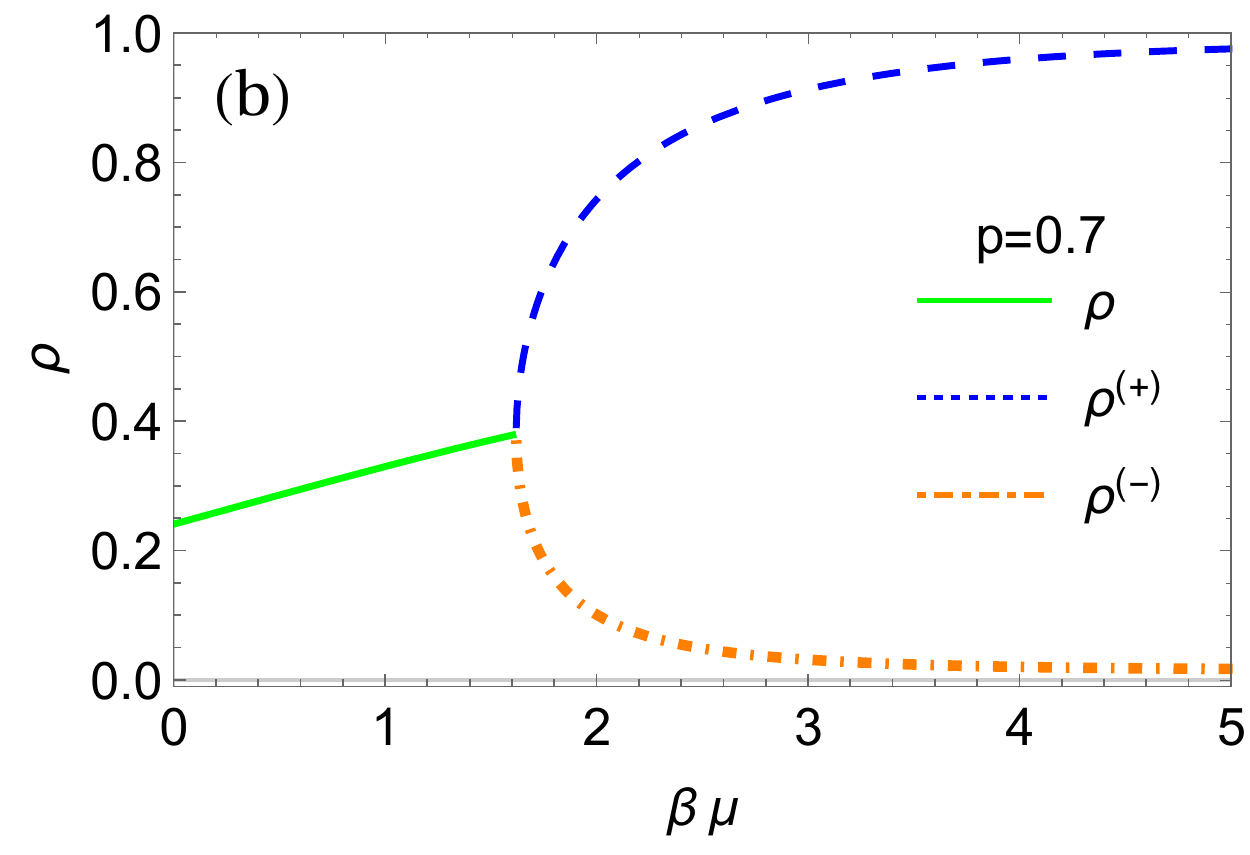}
	\end{center}
	\caption{(Colour online) The Husimi lattice with $J = 0$. The particles' mean density as a function of $\beta \mu$ for two cases of the disordered degree: (a) $p = 0.3<p^{\rm (Hus)}$, (b) $p = 0.7>p^{\rm (Hus)}$. The green curves correspond to density $\rho = \rho^{(A)} = \rho^{(B)}$, blue and orange curves correspond 
		to the particles mean densities in the phase with a broken symmetry, whose limiting values 
		$\rho^{(+)}$ and $\rho^{(-)}$ are given in equations \eqref{pmh}. The transitions from a symmetric phase into the phase with a broken symmetry occurs at the critical values of chemical potential defined in equation (\ref{crit_eq}).}
	\label{densityH}
\end{figure} 

In Fig.~\ref{densityH} we depict the mean densities of the $A$ and $B$ particles as functions of $\beta \mu$  for two cases: $p < p^{\rm (Hus)}$ and $p > p^{\rm (Hus)}$. We observe that, in general, the behaviour is very similar to the one found in case of the Bethe lattice  (see Fig. \ref{fig3}) and differs only in the loci of the critical points. Indeed, we see that for   $p=0.3 < p^{\rm (Hus)}$ both mean densities are equal to each other for any value of $\beta \mu$ and tend towards their limiting value $= 1/2$ when the chemical potential tends to infinity.  For $p=0.7 > p^{\rm (Hus)}$, likewise it happens in case of the Bethe lattice (see Fig. \ref{fig3}b),  the 
mean densities are equal to each other for sufficiently small values of $\beta\mu$ and then, at a certain value of the chemical potential, the symmetry is broken and one the species (which is equally probable for either $A$ or $B$)
gets present in majority with the mean density $\rho^{(+)}$, while the second one becomes a minority component with the mean density $\rho^{(-)}$. In the limit $\beta \mu \to \infty$, the densities in the phase with a broken symmetry approach their limiting $p$-dependent values
	\begin{eqnarray}
	\label{pmh}
		\rho^{(+)} \Big|_{(\beta \mu) \to \infty} &=& \frac{1}{2} + \frac{\sqrt{p (2 - p) \Big(5 (2 - p) p - 4\Big)}}{2 \Big(3 (2 - p) p - 2\Big)}, \nonumber\\
		\rho^{(-)} \Big|_{(\beta \mu) \to \infty} &=& \frac{1}{2} - \frac{\sqrt{p (2 - p) \Big(5 (2 - p) p - 4\Big)}}{2 \Big(3 (2 - p) p - 2\Big)},
	\end{eqnarray}
which have a bit more complicated form than the ones found for the Bethe lattice (see equation \eqref{pm}). Note that similarly to the behaviour on the Bethe lattice, a complete breaking of the symmetry occurs only for $ p \equiv 1$, and in this case
only $\rho^{(+)} = 1$ and $\rho^{(-)} = 0$. 
For any intermediate value of $p$ bounded away from $1$, both the majority and the minority 
components are present in the system. We also note parenthetically that, likewise as it was observed for the Bethe lattice, the values of the  mean densities on the Husimi lattice are not very far from $1/3$ for zero chemical potential  (see  \ref{AppC}). Lastly, in Fig. \ref{fig4} we depict the order parameter  $\Delta\rho$ for the model on the Husimi lattice, which is defined explicitly by (see \ref{B} for the details of a derivation)
\begin{flalign}
\label{orderhus}
\Delta\rho = |\rho^{(A)} {-} \rho^{(B)}| =  \frac{4 z u v}{1 + 2 z \left(u^{2} + v^{2}\right)}.
\end{flalign}
Note that $\Delta\rho$  is equal to zero ($v=0$) within the symmetric phase, and has a non-zero value within the phase with a broken symmetry. As one infers from Fig.~\ref{fig4},  the order parameter on the Husimi lattice shows essentially the same behaviour as
its counterpart for the Bethe lattice. The only differences are a) that the critical point on the Husimi lattice 
is shifted towards smaller values of the chemical potential, and b) that $\Delta\rho$ on the Husimi lattice 
attains substantially higher values for fixed $\beta \mu$ than its counterpart on the Bethe lattice, meaning that the effect of a broken symmetry is higher on the Husimi lattice than on the Bethe one.

Finally, we address a question whether on the Husimi lattice there exist phases with an alternating order, which we have observed for the Bethe lattice. As a matter of fact, here the situation appears to be quite delicate, 
as it was shown in Ref.~\cite{Monroe1998}. The point is that for systems   
with repulsive inter-particle interactions (or antiferromagnetic interactions for spin models)  
defined on lattices 
with a geometrical frustration a robust analysis should be based from the very beginning 
on a description which involves all 
possible sub-lattices. In other words, in our case one has to consider simultaneously 
the recursions obeyed by the auxiliary partition functions on all three sub-lattices, i.e. we have to face six coupled non-linear equations instead of equations \eqref{systeqH}.  In particular, such an approach is the only way to 
determine the 
antiferromagnetic phase transition on the square Husimi lattice \cite{Huang2014}. 
In doing so, we have found by checking obtained equations numerically that for the case under study with $t=2$ the physically plausible expressions for the roots, that have to be real and positive, 
arise only when the auxiliary partition functions defined for different sub-lattices are equal to each other. Recall that an analogous observation was made for the  antiferromagnetic spin-$1$ model defined on the Husimi lattice  \cite{Jurcisinova2016}.
We therefore conclude that on the Husimi lattice with $t = 2$ there is no alternating order phase. Of course, this  does not rule out such a possibility for similarly constructed pseudo-lattices with $t \geq 3$.   We note that 
it appears
technically very difficult to tackle the systems
 with $t \geq 3$, 
in view of a strongly non-linear character of the recursion relations~\eqref{systeqH}. Such an analysis will be presented elsewhere.

%%%%%%%%%%%%%%%%%%%%%%%%%%%%%%%%%%%%%%%%%%%%%%%%%%%%%%%%%%%%%%%%
\section{Conclusions}
\label{conc}
%%%%%%%%%%%%%%%%%%%%%%%%%%%%%%%%%%%%%%%%%%%%%%%%%%%%%%%%%%%%%%%%
To recapitulate, we studied here thermodynamic equilibrium
properties of a binary lattice-gas comprising
interacting $A$ and $B$ particles, (or, in other words, a ternary mixture of two kinds of particles, and voids), 
which undergo continuous exchanges with their respective reservoirs, maintained
at equal chemical potentials, $\mu_A = \mu_B = \mu$. Apart from the hard-core exclusion, particles of similar species 
experience nearest-neighbour interactions of amplitude $J$, which is the same for $AA$ and $BB$ pairs and may be positive or negative. In turn, neighbouring particles of dissimilar species \textit{repel} each other, which represents a rather unusual physical situation.  In our settings such a repulsion between dissimilar species emerges naturally 
within an analysis of the binary reactive lattice gases model in presence of special catalytic bonds, with random annealed spatial  distribution and mean concentration $p$. As a consequence, 
the magnitude of the repulsion is controlled by $p$: $p=0$ corresponds to 
a zero repulsion, while $p=1$ -- to an infinitely strong repulsion, when an $A$ and a $B$ cannot reside simultaneously on two neighbouring sites. For intermediate values of $p$, the repulsive interactions have a finite magnitude. Overall, our model is closely related to well-studied models of 
the so-called hard-objects on regular or pseudo-lattices.

For two kinds of standard pseudo-lattices - the Bethe lattice and the Husimi lattice - we determined the full phase diagram of such a lattice gas. We showed that the latter is rather complicated and contains several phases. More specifically, we demonstrated that there exists a phase with a spontaneously 
broken symmetry between $A$s and $Bs$, in which the species are present at two distinct mean densities, despite the fact that all the parameters are the same for both species. Further on, there is a symmetric phase, in which the species are present at equal mean densities. A transition from a symmetric phase into a phase with a broken symmetric may be of the first order or continuous. 
Such two phases exist on both the Bethe lattice and the Husimi lattice and only the 
precise location of the critical points is somewhat different; we realised that, in general, for the same $J$ and $p$, the symmetry is broken on the Husimi lattice at smaller values of the chemical potential  than on the Bethe one. 

Lastly, we showed that on the Bethe lattice there exist two phases with a structural order: in one of them 
the system spontaneously partitions into two sub-lattices one of which is occupied by both kinds of particles
 present with the same mean density, while the second sub-lattice is almost empty. In the second phase, the system again splits into two sub-lattices 
one of which is occupied by one kind of species, while the second sub-lattice is occupied by the other one. The system enters the first phase and leaves it via a continuous transition, and also enters the second phase via a continuous transition.
Such phases are absent on the Husimi lattice due to stronger frustration effects.

\section*{Acknowledgments}

 M.D. and D.S.  wish to thank Yurij Holovatch and Ihor Mryglod for useful discussions.
 % during webinars of Institute for Condensed Matter Physics. 
 M.D. acknowledges a financial support from the Polish National Agency for Academic Exchange (NAWA) through the Grant No.
PPN/ULM/2019/1/00160, and also a support from the National Academy of Sciences of Ukraine within the framework of the Project K$\Pi$KBK 6541230.

\newpage
%%%%%%%%%%%%%%%%%%%%%%%%%%%%%%%%%%%%%%%%%%%%%%%%%%%%%%%%%%%%%%%%

\appendix

%%%%%%%%%%%%%%%%%%%%%%%%%%%%%%%%%%%%%%%%%%%%%%%%%%%%%%%%%%%%%%%%

\section{The Bethe lattice}
\label{A}

\subsection{Derivation of the recursion relations obeyed by $x_N$ and $y_N$.}
\label{DER}

 We focus on the derivation of the recursion relations \eqref{FinalCayleyRec2}.
 To render our derivation more transparent, we
 let the activities of $A$ and $B$ particles be different, (denoting them as $z_A$ and $z_B$, respectively), and also let the amplitudes of the $A-A$ and $B-B$ interactions be different, $J_A$ and $J_B$.   This will permit us to highlight different contributions in a more explicit way. Moreover, keeping the activities different will permit us to evaluate the mean densities of both species, by a mere
  differentiation of the free energy over the corresponding activity.  In the final results, we will eventually return to the symmetric case.

 The first step consists in considering three possible events with respect to the occupation
 of the root site $O$.
  The grand partition function (\ref{averagedZnCayley}) in this case   can be written  for $N$ generations of entire Cayley tree:
 \begin{eqnarray}
 \label{ZCayley2}
 Z (p) = Z_{N}^{(0)}(p) + Z_{N}^{(A)}(p) + Z_{N}^{(B)}(p),
 \end{eqnarray}
 where $Z_{N}^{(0)}(p)$,  $Z_{N}^{(A)}(p)$, $Z_{N}^{(B)}(p)$ are  grand partition functions with  vacant root site,  occupied  by an $A$ particle and occupied by  a $B$ particle, respectively.
   The Cayley tree with the specified occupation of  $O$ naturally decomposes into
  $q$ independent branches,
  each being a rooted tree with a prescribed occupation of the root site, which leads to
   our equation \eqref{ZCayley}. We introduce then auxiliary functions  $C_{N}(0,p)$, $C_{N}(A,p)$ and $C_{N}(B,p)$ through the relations :
\begin{flalign}
\label{SystCayley}
Z_{N}^{(0)}(p) = C_{N}^{q}(0,p), \quad %\nonumber \\
Z_{N}^{(A)}(p) = z_{A}^{1-q} C_{N}^{q}(A,p),\quad  %\nonumber \\
Z_{N}^{(B)}(p) = z_{B}^{1-q} C_{N}^{q}(B,p).
\end{flalign}
We note that each of the rooted trees contains $q-1$ identical sub-branches (which are also rooted trees) with $N$ generations.  At the second step, we consider all possible values of the occupation variables of the sites neighbouring to the root. In doing so, we realise  that the auxiliary functions in equation \eqref{SystCayley} obey
\begin{eqnarray}
\label{CayleyRecur}
C_{N}(0,p) &=& C_{N-1}^{q-1}(0,p) + z_{A}^{2-q} C_{N-1}^{q-1}(A,p) + z_{B}^{2-q} C_{N-1}^{q-1}(B,p), \nonumber \\
C_{N}(A,p) &=& z_{A} C_{N-1}^{q-1}(0,p) + z_{A}^{3-q} e^{\beta J_A} C_{N-1}^{q-1}(A,p) \nonumber \\ &+& (1-p) z_{A} z_{B}^{2-q} C_{N-1}^{q-1}(B,p), \nonumber \\
C_{N}(B,p) &=& z_{B} C_{N-1}^{q-1}(0,p) + (1-p) z_{A}^{2-q} z_{B} C_{N-1}^{q-1}(A,p) \nonumber \\ &+& z_{B}^{3-q} e^{\beta J_B} C_{N-1}^{q-1}(B,p) ,
\end{eqnarray}
which permit us to eventually establish the desired recursion relations. The latter simplify once we introduce new variables $x_N$ and $y_N$, defined as the ratios
 of the auxiliary functions,
\begin{eqnarray}
\label{newvar}
x_{N} = \frac{C_{N}(A,p)}{z_{A}C_{N}(0,p)}, \qquad {\text{and}} \qquad y_{N} = \frac{C_{N}(B,p)}{z_{B}C_{N}(0,p)}.
\end{eqnarray}
In terms of $x_N$ and $y_N$, we get a system of two coupled non-linear recursion relations:
\begin{eqnarray}
\label{FinalCayleyRec}
x_{N} = \frac{1 + z_{A} e^{\beta J_A} x_{N-1}^{q-1} + (1-p) z_{B} y_{N-1}^{q-1}}{1 + z_{A} x_{N-1}^{q-1} + z_{B} y_{N-1}^{q-1}}, \nonumber \\
y_{N} = \frac{1 + (1-p) z_{A} x_{N-1}^{q-1} + z_{B} e^{\beta J_B} y_{N-1}^{q-1}}{1 + z_{A} x_{N-1}^{q-1} + z_{B} y_{N-1}^{q-1}}.
\end{eqnarray}
In the symmetric case under study here, i.e. when $J_A = J_B = J$ and $z_A = z_B = z$, equations \eqref{FinalCayleyRec} reduce to our equations \eqref{FinalCayleyRec2}.

 Mean  densities of $A$  and $B$ species on the central site  $O$ of the Cayley tree can be expressed in the following form
 \begin{eqnarray}
 \label{meandensZ}
 \rho_{0}^{(A)} {=} \frac{ Z_{N}^{(A)}(p)}{  Z_{N}^{(0)}(p){+}Z_{N}^{(A)}(p){+} Z_{N}^{(B)}(p)}, \quad
 \rho_{0}^{(B)} {=}  \frac{ Z_{N}^{(B)}(p)}{  Z_{N}^{(0)}(p){+}Z_{N}^{(A)}(p){+} Z_{N}^{(B)}(p)}.
 \end{eqnarray}
 Using (\ref{SystCayley}) and (\ref{newvar}) we get expressions  in the variables $x_{N}$ and $y_{N}$:
\begin{eqnarray}
\label{meandens}
\rho_{0}^{(A)} = \frac{z_{A} x_{N}^{q}}{1 + z_{A} x_{N}^{q} + z_{B} y_{N}^{q}}, \qquad
\rho_{0}^{(B)} = \frac{z_{B} y_{N}^{q}}{1 + z_{A} x_{N}^{q} + z_{B} y_{N}^{q}}.
\end{eqnarray}

%%%%%%%%%%%%%%%%%%%%%%%%%%%%%%%%%%%%%%%%%%%%%%%%%%%%%%%%%%%%%%%%
\subsection{Free energy of a binary mixture of particles on the Bethe lattice.}\label{Bfree}
%%%%%%%%%%%%%%%%%%%%%%%%%%%%%%%%%%%%%%%%%%%%%%%%%%%%%%%%%%%%%%%%
Derivation of the free energy of a binary lattice-gas of particles (a ternary mixture of particles and voids) on the Bethe lattice
follows the general approach
of Ref.~\cite{Ananikian1998}, which applies to (rather) arbitrary recursive lattices.
Substituting equations~\eqref{SystCayley} into expression~\eqref{ZCayley}, we find that the free energy of a tree with $N$-generations obeys:
\begin{eqnarray}
\label{freeen1}
-\beta F_{N} = \ln Z_{N} = q \ln C_{N}(0,p) + \ln (1 + z_{A} x_{N}^{q} + z_{B} y_{N}^{q}).
\end{eqnarray}
Taking the advantage of the recursions~\eqref{CayleyRecur}, we have then that
 \begin{equation}
 \label{freee3}
 - \beta F_{N} = - (q-1)^{n} \beta F_{N-n} - \beta F_{N n} \,,
\end{equation}
 where $F_{N n}$ stands for the free energy of the model on a subtree with $n$ generations ($n \leq N$) within the Cayley tree with $N$ generations. The latter property
 is given by:
\begin{align}
\label{zu}
- \beta F_{N n} &{=} q \sum_{K=1}^{n} (q {-} 1)^{K{-}1} \ln{\left(1 {+} z_{A} x_{N {-} K}^{q{-}1} {+} z_{B} y_{N {-} K}^{q{-}1}\right)}  \nonumber \\ & {-} (q {-} 1)^{n} \ln{\left(1 {+} z_{A} x_{N {-} n}^{q} {+} z_{B} y_{N {-} n}^{q}\right)} {+} \ln{\left(1 {+} z_{A} x_{N}^{q} {+} z_{B} y_{N}^{q}\right)}.&&
\end{align}
Further on, in the limit $N \to \infty$ (such that all $x_{N-K} \equiv x$ and $y_{N-K} \equiv y$) the resulting expression for $F_n$ attains the form
\begin{eqnarray}
\label{freeen2}
-\beta F_{n} &=& q \frac{(q-1)^{n}-1}{q-2} \ln (1 + z_{A} x^{q-1} + z_{B} y^{q-1}) \nonumber \\ &-& ((q-1)^{n}-1) \ln (1 + z_{A} x^{q} + z_{B} y^{q}) .
\end{eqnarray}
The last step consists in turning to the limit of the so-called Bethe lattice - a deep interior
of the Cayley tree far away from the boundary sites. Within this interior part, which is the Cayley tree with $n$ generations, all the bulk sites are considered to be equivalent.  According to \cite{baxter2} (see also Refs. \cite{Gujrati1995} and \cite{Ananikian1998} for an additional discussion), the number
  $N^B_{s}$ of such sites 
   is simply related to the number  $N^B_{b}$ of bonds via the homogeneity assumption,
$N^B_{b}/N^B_{s} = q/2$.  For the Cayley tree with $n$ generations one has
$N^B_{b} = q((q - 1)^n - 1)/(q-2)$.
 % Note that, in general,
%the effect of the boundary sites is not at all negligible - for the Cayley tree both the number of the interior sites and
%the number of the boundary sites
%increase at exactly the same rate which us proportional  to $(q-1)^{N}$ \cite{baxter2}. 
%In consequence, their respective contributions to the partition function have the same order and even partially compensate each other suppressing a critical behaviour.  
 Therefore, the number of sites is given by $N^B_{s} = 2 ((q-1)^{n}-1)/(q-2)$, such that
we get:
\begin{eqnarray}
\label{FreeEnergy}
-\beta f &=& - \frac{\beta F_{n}}{N^B_{s}} = \frac{q}{2} \ln (1 + z_{A} x^{q-1} + z_{B} y^{q-1}) \nonumber \\  &-& \frac{q-2}{2} \ln (1 + z_{A} x^{q} + z_{B} y^{q}) .
\end{eqnarray}
This is the desired expression for the free energy on the Bethe lattice.

%%%%%%%%%%%%%%%%%%%%%%%%%%%%%%%%%%%%%%%%%%%%%%%%%%%%%%%%%%%%%%%%
\subsection{Fixed point solutions of recursions}
%%%%%%%%%%%%%%%%%%%%%%%%%%%%%%%%%%%%%%%%%%%%%%%%%%%%%%%%%%%%%%%%

{\em Symmetric phase versus the phase with a broken symmetry.}
Let us suppose from now on that the strength of interactions between similar species is the same, i.e.,  $J_{A}=J_B=J$. Then, we assume that in the limit $N \to \infty$ the auxiliary variables $x_N \to x$ and $y_N \to y$, such that in this limit the system of equations \eqref{FinalCayleyRec} takes the form:
\begin{eqnarray}
x = \tilde g(x,y),\qquad \tilde g(x,y)= \frac{1 +  e^{\beta J}\, z_A x^{q-1} + (1-p)z_B y^{q-1}}{1 + z_A x^{q-1} + z_B y^{q-1}},\label{eqx}\\
y = \tilde h(x,y), \qquad \tilde h(x,y)= \frac{1 + (1-p)\,z_A x^{q-1} + e^{\beta J} z_B y^{q-1}}{1 + z_A x^{q-1} + z_B y^{q-1}} \label{eqy}.
\end{eqnarray}
Free energy per site of the Bethe lattice then obeys equation \eqref{FreeEnergy}.

Since we consider an infinitely deep interior of the Cayley tree, in which all sites are equivalent, mean densities of both kinds of particles are defined by expressions \eqref{meandens}, 
\begin{eqnarray}
\label{meandens2}
\rho^{(A)} = \frac{z_{A} x^{q}}{1 + z_{A} x^{q} + z_{B} y^{q}}, \qquad
\rho^{(B)} = \frac{z_{B} y^{q}}{1 + z_{A} x^{q} + z_{B} y^{q}}.
\end{eqnarray}
Note that our equation \eqref{meandens4} in the main text follows from the latter expression by setting $z_A = z_B = z$.

We thus now have all necessary ingredients for our analysis.
Following Ref. \cite{Ananikian1991}, we first formally cast equations \eqref{eqx} and \eqref{eqy}
 into the form
\begin{eqnarray}
z_A x^{q-1} =  \frac{x(1-e^{\beta J})-p y +p+e^{\beta J}-1}{(1 -p -  e^{\beta J})(1 -p +e^{\beta J} -x-y)}\label{eqx2}\\
z_B y^{q-1}  = \frac{y(1-e^{\beta J})-p x +p+e^{\beta J}-1}{(1 -p -  e^{\beta J})(1 -p +e^{\beta J} -x-y)} \label{eqy2}.
\end{eqnarray}
Next, introducing variables $u=(x + y)/2$ and $v=(x - y)/2$, we get two following equations
\begin{eqnarray}
\label{zazb}
z_A z_B  =  \frac{((u-1)^2-\alpha^2v^2)}{4(\gamma-u)^2(u^2-v^2)^{q-1}} , \label{eqx3}\\
z_A/z_B   = \frac{(u-1+\alpha v)(u-v)^{q-1}}{(u-1-\alpha v)(u+v)^{q-1}},  \label{eqy3}
\end{eqnarray}
where  $\alpha$ and $\gamma$ are given by
\begin{equation}\label{param}
\alpha=\frac{1-e^{\beta J}+p}{1-e^{\beta J}-p}, \qquad \gamma=(1-p+e^{\beta J})/2 \,.
\end{equation}
Standard analysis (see Ref. \cite{Ananikian1991} for more details), which we perform only for the symmetric case $z_A=z_B = z$, then shows that
 there are two different fixed point solutions corresponding to two
stable thermodynamics phases:\\
(i) The solution with $ v = 0 $ is given by
	\begin{eqnarray}
	\label{dis1a}
	z=\frac{u-1}{2(\gamma-u)u^{q-1}},
	\end{eqnarray}
which corresponds to a situation with equal mean densities of $A$ and $B$ species.
This solution thus describes the disordered symmetric phase.\\
(ii) The solution with $v \neq 0$ is defined by the two following equations	
\begin{flalign}
	\label{ord1.1a}
	\left\{\begin{array}{ c }
	(\alpha-(q-1)\frac{u-1}{u})+(C^2_{q-1}\alpha-C^2_{q-1}\frac{u-1}{u})\left(\frac{v}{u}\right)^2+\dots+\alpha\left(\frac{v}{u}\right)^{q-1}\\
	\mbox{for}\quad q\quad \mbox{ odd}\\
	(\alpha{-}(q{-1})\frac{u{-}1}{u}){+}(C^2_{q-1}\alpha{-}C^2_{q-1}\frac{u-1}{u})\left(\frac{v}{u}\right)^2{+}{\dots}{+}(\alpha(q-1){-}\frac{u-1}{u})\left(\frac{v}{u}\right)^{q-1}\\
		\mbox{for}\quad q\quad \mbox{ even}
	\end{array}\right\}{=} 0, \nonumber\\
	\end{flalign}
where $C^k_n$ stands for the binomial coefficient, and
	\begin{flalign}
	\label{ord1.2a}
z^2=\frac{(u-1)^2-\alpha^2 v^2}{4(\gamma-u)^2 (u^2-v^2)^{q-1}}  \,.
	\end{flalign}
This solution defines the phase with a spontaneously broken symmetry between the species.  Correspondingly, the demarkation surface between the symmetric phase and the phase with a broken symmetry in the region of continuous transitions obtains by substituting $v=0$ into the expressions \eqref{dis1a}, which yields $u_c=(q-1)/(q-1-\alpha)$ at the transition. Then substituting the latter expression for $u_c$  into  equation \eqref{ord1.1a} and setting 
$v=0$, we get
\begin{eqnarray}
	\label{dis1}
	z_c=\frac{u_c-1}{2(\gamma-u_c)u_c^{q-1}} .
	\end{eqnarray}
 In turn, the line of the trictitical points is obtained by using the condition
	 \begin{eqnarray} \label{tricrit}
	\frac{\delta z}{\delta u} \Big|_{v=0} = \frac{\partial z}{\partial u} \Big|_{v=0} +\frac{\partial z}{\partial v^2} \frac{\partial v^2}{\partial u}\Big|_{v=0}  =0 \,,
	\end{eqnarray} which is to be taken at $u=u_c$. Substituting  equation \eqref{ord1.2a} into condition \eqref{tricrit}, we get
 \begin{eqnarray}\label{tric}
	\frac{(q-1)}{(\gamma - 1)(q-1-\alpha)-\alpha} -q+2-\frac{(q-3)(q-1-\alpha)}{2q (q-1)}=0.
	\end{eqnarray}	
	The line of first order phase transitions, which meets at the tricritical fixed point with the line of second order phase transitions, can be obtained by equating the free energies (see equation \eqref{FreeEnergy}) calculated for different phases.
	
Lastly, we note that the obtained expressions for the limiting values of auxiliary variables $x$ and $y$, together with the expressions \eqref{meandens2} which defines  the mean densities, permit us to introduce a natural order parameter $\Delta \rho =  |\rho^{(A)} - \rho^{(B)}| $. The latter  is given explicitly by
\begin{equation}
\label{OP}
\Delta \rho = v \frac{u (1 + \alpha) - 1}{u (u - 2) + \alpha v^{2} + \gamma} \,.
\end{equation}
This parameter is exactly equal to zero within the random symmetric phase and is non-zero within the phase with a broken symmetry (see Fig. \ref{fig4}). In \ref{scaling} below, we analyse its behaviour for $\mu$ close to the critical value $\mu_c$ of the chemical potential.

{\em Symmetric phase with a structural order.}	
As we have already mentioned in the main text, equations describing
the symmetric phase with an alternating order are obtained from the recursion scheme
	\eqref{FinalCayleyRec} by re-iterating it once more, in order to express $x_N$ and $y_N$ through $x_{N-2}$ and $y_{N-2}$, in which $N$ and $N-2$ have the same parity.  Turning to the limit $N \to \infty$, we formally rewrite equations
	\eqref{eqx} and \eqref{eqy} as
	\begin{eqnarray}
	\label{a}
	x=\tilde g(\tilde g(x,y),\tilde h(x,y)) \,, \quad y = \tilde h(\tilde g(x,y),\tilde h(x,y)) \,.
	\end{eqnarray}
 We find then that for $q = 3$, for some negative $J$ and for a certain interval of values of $p$ 
 equations \eqref{a} possess three  kinds of solutions with $x = y$ (or $v =0$): \\
-- Solution $x_{odd} = x_{even} = y_{odd} = y_{even}$,  equation \eqref{dis1a}, which corresponds to a situation with equal mean densities of $A$ and $B$ particles and no alternating order.  \\
-- Solution $x_{odd} = y_{odd} \neq x_{even} = y_{even}$ is one of the solutions of
the following quadratic equation, obtained by re-iterating the recursion scheme,  
\begin{equation}
	\label{f2J}
 u^2 z \left(z \left(e^J-p+1\right)^2+2\right)+u z \left(e^J-p-1\right)+z \left(e^J-p+1\right)+1 = 0, \,\,\, u = (x + y) / 2 \,.
\end{equation}
Note that here there is an alternating order, when the system spontaneously partitions into two different sub-lattices:  
 $A$s and $B$s occupy predominantly one of them, while the second one is almost empty. This phase (PAO I) is defined by 
the following inequality 
\begin{equation}
\label{discriminant}
z \left(1+ p -e^{\beta J} \right)^2-4 \Big(z \left(1 - p + e^{\beta J}\right)+1\Big) \left(z \left(1 - p + e^{\beta J}\right)^2+2\right)  \geq 0 ,
\end{equation}
which gives, in turn, the condition that the solutions of equation \eqref{f2J} are real.  \\
-- Solution $x_{odd} = y_{even} \neq x_{even} = y_{odd}$. We find this solution from the following quadratic equation:
\begin{flalign}
	\label{f3J}
	&z \left(u_{1}^{2} + e^{2\beta J}\right) \left(e^{\beta J} + p - 1\right) + \left(u_{1} - e^{\beta J}\right) \sqrt{z \left(e^{\beta J} + p - 1\right)} \left(\left(u_{1} + p - 1\right) \sqrt{z \left(e^{\beta J} + p - 1\right)} \right.&& \nonumber \\ & - \left. \sqrt{z \left(e^{\beta J} - p + 1\right)^{2} \left(e^{\beta J} + p - 1\right) + 4 \left(e^{\beta J} - p - 1\right)}\right) + 2 e^{\beta J} - 2 = 0,&&
\end{flalign}
where $u_{1} = (x_{odd} + y_{even}) / 2$,  $u_{2} = (x_{even} + y_{odd}) / 2$ and hence,
\begin{equation}
	\label{u2}
	u_{2}^{2} = \frac{u_{1} - 1 + z \, u_{1}^{2} \left(u_{1} - 1 + p\right)}{z \left(e^{\beta J} - u_{1}\right)}.
\end{equation}
Solutions of equations \eqref{f3J}) and \eqref{u2} correspond to another kind of an alternating order, when the system spontaneously partitions into two different sub-lattices containing predominantly either kind of particles, i.e. $A$s and $B$s occupy predominantly different sub-lattices  (PAO II). This phase is defined by the inequality:
\begin{equation}
	\label{discriminantII}
	4 \, z \, (3 e^{\beta  J}+p-1) (e^{\beta  J}+p-1)^2+(3e^{\beta  J}+p-3)^2 \geq 0 .
\end{equation}

%%%%%%%%%%%%%%%%%%%%%%%%%%%%%%%%%%%%%%%%%%%%%%%%%%%%%%%%%%%%%%
\subsection{Critical exponents \label{scaling}}
%%%%%%%%%%%%%%%%%%%%%%%%%%%%%%%%%%%%%%%%%%%%%%%%%%%%%%%%%%%%%%%%

We focus here on the critical exponent describing scaling behaviour of the order parameter in the vicinity of a critical point.
Following Ref. \cite{Ananikian1991} let us denote
\begin{eqnarray}
u - u_{c} = u_{c} \delta, \quad v = u_{c} \varepsilon,
\end{eqnarray}
where $u_{c} = (q - 1)/(q - 1 - \alpha)$. Expanding the expressions \eqref{ord1.1a} and \eqref{ord1.2a} near the line of critical points, we obtain
\begin{flalign}
{h}& { = \varepsilon \left[1 - q + \alpha a + \left(q - 1 - \alpha a^{2}\right)\delta  + \frac{1}{3}\left(1 - q  + \alpha^{3} a^{3}\right)\varepsilon^{2} + \left(1 - q + \alpha a^{3}\right)\delta^{2} \right.}&& \nonumber \\
&{\left.+ \left(q - 1 - \alpha^{3} a^{4}\right)\delta \varepsilon^{2} + \frac{1}{5} \left(1 - q + \alpha^{5} a^{5}\right)\varepsilon^{4} + \mathcal{O}\left(\varepsilon^{6}, \varepsilon^{4} \delta, \varepsilon^{2} \delta^{2}, \delta^{3}\right) \right]}, &&\\	
\label{muexp}
\mu - \mu_{c} &= -\left(q - 1 - a - b\right)\delta + \frac{1}{2}\left(q - 1 - \alpha^{2} a^{2} \right)\varepsilon^{2} + \frac{1}{2} \left(q - 1 - a^{2} + b^{2}\right) \delta^{2}&& \nonumber \\
&- \left(q - 1 - \alpha^{2} a^{3}\right)\varepsilon^{2} \delta + \frac{1}{4} \left(q - 1 - \alpha^{4} a^{4}\right)\varepsilon^{4} + \mathcal{O}\left(\varepsilon^{6}, \varepsilon^{4} \delta, \varepsilon^{2} \delta^{2}, \delta^{3}\right),&&
\end{flalign}
where {$h=(\mu_A-\mu_B)/2$, $\mu=(\mu_A+\mu_B)/2$}, $\mu_c$ is defined in equation \eqref{dis1}, while $a = u_{c}/(u_{c}-1)$ and $b = u_{c}/(\gamma - u_{c})$. Consequently, the behaviour of the order parameter (defined  in equation \eqref{OP}) in a vicinity
of the critical point is given by
\begin{eqnarray}\label{drho}
\Delta \rho = \frac{u_{c} \left[u_{c}(1 + \alpha) - 1\right]}{u_{c}(u_{c} - 2) + \gamma} \varepsilon + \mathcal{O}\left(\delta,\varepsilon^{3}\right).
\end{eqnarray}
Next, using the condition $h= 0$, we find that $\delta = \varepsilon^2 (q - 2) q (u_c - 1)/3$ at the critical points, and hence, it follows from equation \eqref{muexp} that
\begin{eqnarray}
\label{zi}
\mu - \mu_{c} = \mu_{1} \Delta \rho^{2} + \mathcal{O}\left(\Delta \rho^{4}\right),
\end{eqnarray}
with
\begin{flalign}
\mu_{1} &= \frac{(q-2)}{3q^2}\frac{{u_{c}(u_{c} - 2) + \gamma}}{(u_c-1)^2u_c^3}\left(\frac{(q-1)}{{(\gamma - 1)}(q-1-\alpha)-\alpha} -q+2-\frac{(q-3)(q-1-\alpha)}{2q (q-1)}\right)&&
\end{flalign}
It follows thus that in the vicinity of a line of critical points a scaling behaviour of the order parameter is characterised by a classical mean-field critical exponent $\beta=1/2$ (i.e. $\Delta \rho\sim|\mu-\mu_c|^\beta$). This is not true, however, for the tricritical point at which $\mu_1$ vanishes, $\mu_1=0$. At the tricritical point the expansion \eqref{zi} ensures that the  tricritcal exponent $\beta_2 =1/4$.

%%%%%%%%%%%%%%%%%%%%%%%%%%%%%%%%%%%%%%%%%%%%%%%%%%%%%%%%%%%%%%%%

\section{The Husimi lattice}
\label{B}

\subsection{Recursions obeyed by $x_N$ and $y_N$.}

 The grand canonical partition function on the  Husimi lattice with $N$ generations is first written in the same form as the one for the Bethe lattice (see equation \eqref{ZCayley2}),  by considering three possible events with respect to the occupation of the root site. Then, 
 expressing  the grand canonical partition functions of the rooted trees with a specified occupation of the root site $O$, i.e. 
$Z_{N}^{(0)}(p)$,  $Z_{N}^{(A)}(p)$ and $Z_{N}^{(B)}(p)$, respectively, through the  auxiliary functions  $D_{N}(0,p)$, $D_{N}(A,p)$ and $D_{N}(B,p)$:
\begin{flalign}
\label{SystHusimi}
Z_{N}^{(0)}(p) = D_{N}^{t}(0,p), \quad %\nonumber \\
Z_{N}^{(A)}(p) = z_{A}^{1-t} D_{N}^{t}(A,p),\quad  %\nonumber \\
Z_{N}^{(B)}(p) = z_{B}^{1-t} D_{N}^{t}(B,p) \,,
\end{flalign}
we find, following essentially the same procedure as in case of the Bethe lattice, that 
the functions $ D_{N} $ obey recursions of the following form:
\begin{flalign}
%\label{B100}
D_{N}(0,p)&{=} D_{N-1}^{2(t-1)}(0,p) {+} 2z^{2-t}_{A}  D_{N-1}^{t-1}(0,p)D_{N-1}^{t-1}(A,p)   \nonumber \\&{+}2z^{2{-}t}_{B}  D_{N{-}1}^{t{-}1}(0,p)D_{N{-}1}^{t{-}1}(B,p) {+}2z^{2{-}t}_{A} z^{2{-}t}_{B} (1{-}p) D_{N-1}^{t-1}(A,p)D_{N-1}^{t-1}(B,p)  \nonumber \\&{+} z^{2(2{-}t)}_{A} e^{\beta J_A}  D_{N-1}^{2(t{-}1)}(A,p) {+} z^{2(2{-}t)}_{B} e^{\beta J_B}  D_{N-1}^{2(t{-}1)}(B,p),\label{d0}  \\ 
%\end{flalign}
%\begin{flalign}
D_{N}(A,p)&= z_A D_{N-1}^{2(t-1)}(0,p) {+} 2z^{3-t}_{A}  e^{\beta J_A} D_{N-1}^{t-1}(0,p)D_{N-1}^{t-1}(A,p)  \nonumber \\&{+}2z_Az^{2{-}t}_{B} (1-p) D_{N{-}1}^{t{-}1}(0,p)D_{N{-}1}^{t{-}1}(B,p) \nonumber \\& {+}2z^{3{-}t}_{A} z^{2{-}t}_{B} (1{-}p) e^{\beta J_A} D_{N-1}^{t-1}(A,p)D_{N-1}^{t-1}(B,p)  \nonumber \\&{+} z^{5{-}2t}_{A} e^{3\beta J_A}  D_{N-1}^{2(t{-}1)}(A,p) {+} z_A z^{2(2{-}t)}_{B} e^{\beta J_B} (1{-}p)^2 D_{N-1}^{2(t{-}1)}(B,p). \label{dA}
\end{flalign}
An analogous expression for $D_N(B)$ obtains from equation \eqref{dA} by a mere interchange of symbols $A\leftrightarrow B$.

%%%%%%%%%%%%%%%%%%%%%%%%%%%%%%%%%%%%%%%%%%%%%%%%%%%%%%%%%%%%%%%%
\subsection{Free energy and mean densities}
%%%%%%%%%%%%%%%%%%%%%%%%%%%%%%%%%%%%%%%%%%%%%%%%%%%%%%%%%%%%%%%%

Performing the same procedure as it was done in the \ref{Bfree} for the Bethe lattice (see  also Ref.~\cite{Ananikian1998}),  we find that  for the Husimi lattice with $n$ generations which are deeply inside  the tree, the free energy obeys
\begin{flalign}
\label{freeenH}
-\beta F_{n} &= t \frac{2^n(t-1)^{n}-1}{2(t-1)-1} \ln \left({1 + 2 z_{A} x^{t-1} + 2 z_{B} y^{t-1} {+} 2 z_{A} z_{B} (1 - p) x^{t-1} y^{t-1}}\right.&&  \nonumber \\ &\left.{+ z_{A}^{2} e^{\beta J_{A}} x^{2(t-1)} + z_{B}^{2} e^{\beta J_{B}} y^{2(t-1)}}\right) - \left(2^{n}(t-1)^{n}-1\right) \ln (1 + z_{A} x^{t} + z_{B} y^{t}).&&
\end{flalign}
where $x$ and $y$ are the fixed point solution  given by equation \eqref{newvar_h}. According to Ref.~\cite{Ananikian1998}, 
the free energy per site in the bulk of the Husimi tree obtains by 
dividing the expression \eqref{freeenH} by the number of bulk sites $N^H_{s}$, which equals $3 ((t-1)^{n}2^n-1)/(2(t-1)-1)$. Hence,  the free energy per site on the Husimi lattice is given by
\begin{flalign}
	\label{FreeEnergyH}
\hspace{1cm}	-\beta f &= - \frac{\beta F_{n}}{N^H_{s}} = \frac{t}{3} \ln \left({1 + 2 z_{A} x^{t-1} + 2 z_{B} y^{t-1} {+} 2 z_{A} z_{B} (1 - p) x^{t-1} y^{t-1}}\right.&&  \nonumber \\ &\left.{+ z_{A}^{2} e^{\beta J_{A}} x^{2(t-1)} + z_{B}^{2} e^{\beta J_{B}} y^{2(t-1)}}\right) - \frac{2t-3}{3} \ln (1 + z_{A} x^{t} + z_{B} y^{t}).&&
\end{flalign}
We restrict our analysis in the main text to a particular choice $t=2$. For such a choice the number  ($N^H_s$) of bulk sites of the  Husimi tree and the number ($N^B_s$) of bulk sites of the Cayley  tree with $q = 3$ obey $N^B_s/N^H_s=2/3$.

In turn, the mean densities of both kinds of particles at the root site are defined in the general form by equation \eqref{meandensZ}.  Then, using equations \eqref{SystHusimi} and \eqref{newvar_h},  we get the following expression
\begin{eqnarray}
\label{meandens_h}
\rho_{0}^{(A)} = \frac{z_{A} x_{N}^{t}}{1 + z_{A} x_{N}^{t} + z_{B} y_{N}^{t}}, \qquad
\rho_{0}^{(B)} = \frac{z_{B} y_{N}^{t}}{1 + z_{A} x_{N}^{t} + z_{B} y_{N}^{t}}.
\end{eqnarray}
Correspondingly, turning to the fixed point solutions of the recursions, we have that 
the mean densities of the species on the Husimi lattice are given by
\begin{eqnarray}
\label{meandens_h2}
\rho^{(A)} = \frac{z_{A} x^{t}}{1 + z_{A} x^{t} + z_{B} y^{t}}, \qquad
\rho^{(B)} = \frac{z_{B} y^{t}}{1 + z_{A} x^{t} + z_{B} y^{t}}.
\end{eqnarray}

Similarly as in case of Bethe lattice  we can define here  an order parameter as the difference of $\rho^{(A)}$ and $\rho^{(B)}$: $\Delta \rho = |\rho^{(A)} - \rho^{(B)}|$. Using equation (\ref{meandens_h2}), we can find an explicit form of $\Delta \rho$ in the terms of symmetric variables $u = (x+y)/2$ and $v = (x-y)/2$. Finally, the difference of $\rho^{(A)}$ and $\rho^{(B)}$ is given by:
\begin{eqnarray}
\Delta \rho = |\rho^{(A)} - \rho^{(B)}| = \frac{4z u v}{1 + 2 z (u^2 + v^2 )}.
\end{eqnarray}

%%%%%%%%%%%%%%%%%%%%%%%%%%%%%%%%%%%%%%%%%%%%%%%%%%%%%%%%%%%%%%%%
\subsection{The surface of critical points}
%%%%%%%%%%%%%%%%%%%%%%%%%%%%%%%%%%%%%%%%%%%%%%%%%%%%%%%%%%%%%%%%

 In the main text we have shown that for $J > 0$ there are two different kinds of fixed point solutions of the recursion relations \eqref{symm_sysJ}, which correspond to the symmetric phase (see equation \eqref{disH}) and the phase with a broken symmetry  (see equations \eqref{ordH} and \eqref{phase2J_2}), respectively. In order to evaluate the expression which defines implicitly  the surface of critical points, we solve the system of equations \eqref{disH} and \eqref{ordH}. Omitting the intermediate steps,  we 
present below such an equation for $t=2$, $J_{A} = J_{B} = J$ and  $z_{A} = z_{B} = z$:
\begin{flalign}
	\label{critical_eqJ}
& 4 z^{2} e^{8 \beta J} \left(e^{\beta J} {-} 1\right) 
{+} 4 z e^{7 \beta  J} \Bigl(1{-}(1{-}p) (9{-}7 p) z\Bigr) 
{-}e^{6 \beta  J} \Bigl(1{+}4 (3 {+} p) z{-}52 (1{-}p)^2 z^2\Bigr) && \nonumber \\
&{-} 4 z e^{5 \beta  J} \Bigl(8 {-} 10 (2{-}p) p {+} (1{-}p)^3(15{-}11 p) z\Bigr) 
{+} 2 e^{4 \beta  J} \Bigl(3{-}5 (2{-}p) p {+} 52 z {-} 4 p \left(23{-}p (9 {+} p)\right) z&& \nonumber \\
& {-} 78 (1{-}p)^4 z^2\Bigr){-}4 e^{3 \beta  J}\Bigl(1 {-} 5 p {-}  (1{-}p)^2\left(1{-}25 (2{-}p) p\right) z {-} 5 (1{-}p)^5(5 {+} p) z^2\Bigr) && \nonumber \\
&{-}e^{2 \beta  J}  \Bigl(\left(3 {-} 5 (2{-}p) p\right)^2 {+} 4  (1{-}p)^2  \left(55 {-} 121 p {+} 5 p^2(13{-}3 p) \right)z{+}20 (1{-}p)^6 z^2\Bigr)&& \nonumber \\
&{+} 4 (1{-}p) e^{\beta  J} \Bigl(3{-}22 p{+}25 p^2{+}54 (1{-}p)^3 z\Bigr) {-} 4 (1{-}p)^2 \Bigl(1{-}8 p{+}16 z (1{-}p)^2\Bigr) = 0.&&
\end{flalign}
Analysing equation \eqref{critical_eqJ}, we infer that the continuous transition takes place for the following value of the activity $z$:
\begin{eqnarray}
	\label{critical_activity__huisimi}
	z_{c} = \exp{((\beta \mu)_{c})} = \frac{\mathcal{D}-c_{1}}{2 c_{2}},
\end{eqnarray}
where $\mathcal{D}$, $c_1$ and $c_2$ are given by
\begin{flalign}
	\mathcal{D} & = \Bigl(e^{6\beta J} {+} e^{2\beta J} (1{-}p) \left(6 e^{\beta J}{+}15{-}33 p{+}5 (3{-}p) p^2\right) {-}2 e^{4\beta J} \left(4{-}3 (2{-}p) p\right)&& \nonumber \\
	&+ 2 (1{-}p)^3 (4{-}11 e^{\beta J}) \Bigr)\Bigl(e^{3\beta J}+(4-5 e^{\beta J}) (1-p)^2\Bigr)^{1/2},&& \nonumber \\
	c_{1} & =  e^{6\beta J} \left(e^{\beta J}{-}3 {-} p\right) {-} 2 e^{5\beta J} \Bigl(4{-}5 (2{-}p) p\Bigr) {-}2 e^{4\beta J} (1{-}p) \Bigl(13{-}(10{+}p)p\Bigr) && \nonumber \\
	&{+}e^{2\beta J} (1{-}p)^2\left(e^{\beta J} \Bigl(1{-}25 (2{-}p) p\Bigr){-}\Bigl(55{-}\Bigl(121{-}5 p (13{-}3 p)p\Bigr)\Bigr){-}2 (1{-}p)^4\left(8{-}27  e^{\beta J}\right)\right), && \nonumber \\
	c_{2} &= e^{2\beta J} \Bigl(e^{2\beta J}{-}5 (1{-}p)^2\Bigr) \left(e^{\beta J} {-} 1 {+} p\right)^2 \Bigl(e^{2\beta J} \left(e^{\beta J} {+} 1 {-} 2 p\right){-}e^{\beta J} (1{-}p) (3 {+} p) {+} (1{-}p)^2\Bigr).\nonumber&&
\end{flalign}
Equating $c_2$ to zero, which enter equation \eqref{critical_activity__huisimi}, we readily find the threshold value $J(p)$ in equation \eqref{Jph}.

%%%%%%%%%%%%%%%%%%%%%%%%%%%%%%%%%%%%%%%%%%%%%%%%%%%%%%%%%%%%%%%%
\subsection{Line of tricritical points}
%%%%%%%%%%%%%%%%%%%%%%%%%%%%%%%%%%%%%%%%%%%%%%%%%%%%%%%%%%%%%%%%

Here we present the exact equation, defining implicitly the line of tricritical points (see  Fig.~\ref{phdh}) on the Husimi lattice, which is obtained exactly along the same lines as in case of the Bethe lattice (see the \ref{A}). For the Husimi lattice it has a rather complicated form :
\begin{flalign}
	\label{tricritical_eq__husimi}
&e^{12(\beta J)_{tc}}
{+}4 e^{11(\beta J)_{tc}} (1{-}p)
{-}4 e^{10(\beta J)_{tc}} \Bigl(6{-}(2{-}p) p\Bigr)
{-}4 e^{9(\beta J)_{tc}} (1{-}p) \Bigl(26{-}9 (2{-}p) p\Bigr)&&\nonumber \\	
&{+}2 e^{8(\beta J)_{tc}} \Bigl(71{-}16 p {-}  \left(56{-}17 (4{-}p) p\right)p^{2}\Bigr)
{+}4 e^{7(\beta J)_{tc}} \Bigl(168{-}474 p{+}511 p^2{-}\left(281{-}15 (5{-}p) p\right) p^3\Bigr)&&\nonumber \\
&-4 e^{6(\beta J)_{tc}} (1-p)\Bigl(145+113 p-477 p^2 + 5 \left(83{-}7 (5{-}p) p\right) p^3\Bigr)	&&\nonumber \\
&-4 e^{5(\beta J)_{tc}} (1-p)^2\Bigl(446-936 p+383 p^2+\left(39-25 (5-p) p\right) p^3\Bigr)	&&\nonumber \\
&+e^{4(\beta J)_{tc}} (1-p)^2\Bigl(1785-78 p-4357 p^2+3308 p^3-25 \left(29+(6-p) p\right) p^4\Bigr)&&\nonumber \\
&+4 e^{3(\beta J)_{tc}} (1-p)^3\Bigl(391-1264 p + 556 p^2+5 \left(44-19 p\right) p^{3}\Bigr)&&\nonumber \\
&{-}4 e^{2(\beta J)_{tc}} (1{-}p)^4\Bigl(603{-}170 p{-}\left(517{-}52 p\right) p^2\Bigr) {+} 16 (1{-}p)^5 \Bigl(14 e^{(\beta J)_{tc}} (3 {+} 5 p){+}4 (1{-}p)\Bigr) {=}0.&&
\end{flalign}
% % % % % % % % % % % % % % % % %
\section{Mean densities at zero chemical potential and $J=0$ \label{AppC}}

\subsection { The mean  particles'  densities on the Bethe lattice at $\mu = 0$ and $J = 0$.}	In virtue of equation  \eqref{meandens2},  the mean densities of particles of both kinds 
at zero chemical potential and at zero interaction strength $J$ obey
\begin{equation}
	\label{meandens_atzero}
	\rho = \rho_{A} = \rho_{B} = \frac{x^q}{1 + 2 \, x^q},
\end{equation}
where $x$ satisfies the equation 
\begin{equation}
\label{bethe_sol}
	(x - 1) (1 + 2 x^{q-1}) + p \, x^{q-1} = 0.
\end{equation}
For $p = 0$, (i.e., no other interactions between the dissimilar species apart from the hard-core ones),  the only 
real  positive solution of equation \eqref{bethe_sol}  with arbitrary $q$ is $x \equiv 1$. In consequence, equation \eqref{meandens_atzero} entails $\rho \equiv 1/3$. This result is obvious, since in this case we have a three-state model without interactions and external fields; therefore, 
each of the species as well as the vacant sites are present at equal mean densities.   

 For $p\not =0$, the solution of  equation \eqref{bethe_sol} depends on both $q$ and $p$, and the mean density defined by equation \eqref{meandens_atzero} is evidently a decreasing function of these parameters. 
 In Fig.~\ref{fig3} (a), for $q = 3$ we have chosen $p = 0.3$, and thus we find that the density at $\mu = 0$ and $J = 0$ is $\rho \approx 0.2993$, while for $p = 0.7$ in Fig.~\ref{fig3} (b) the density at $\mu = 0$ and $J = 0$ is $\rho \approx 0.2543$.

\subsection{ The mean particles' densities on the Husimi lattice at $\mu = 0$ and $J = 0$.}	Similarly, on the Husimi lattice the mean particles' densities for $\mu = 0$ and $J = 0$ obey 
	\begin{equation}
		\label{h-meandens_atzero}
		\rho = \rho_{A} = \rho_{B} = \frac{x^t}{1 + 2 \, x^t},
	\end{equation}
	where $x$ satisfies the equation
	\begin{equation}
		\label{husimi_sol}
		(x - 1) \left(1 + 4 x^{t-1} (1 + x^{t-1})\right) + p \, x^{t-1} \left(2 + (4 - 3 p) x^{t-1}\right) = 0.
	\end{equation}
For $p = 0$, the only real  positive solution of equation \eqref{husimi_sol} is $x \equiv 1$  for arbitrary $t$, which 
again entails a trivial result $\rho \equiv 1/3$. For $p\not= 0$, the mean density $\rho$ is again a decreasing function of the parameters $p$ and $t$. 
 In Fig.~\ref{densityH} (a), for $t = 2$ we have chosen $p = 0.3$, and thus we find that the density at $\mu = 0$ and $J = 0$ is $\rho \approx 0.2922$, while for $p = 0.7$ in Fig.~\ref{densityH} (b) the density at $\mu = 0$ and $J = 0$ is $\rho \approx 0.2529$, which is somewhat greater than the corresponding value of the mean density on the Bethe lattice.

\newpage
%%%%%%%%%%%%%%%%%%%%%%%%%%%%%%%%%%%%%%%%%%%%%%%%%%%%%%%%%%%%%%
%References
%%%%%%%%%%%%%%%%%%%%%%%%%%%%%%%%%%%%%%%%%%%%%%%%%%%%%%%%%%%%%%

\end{document}